\documentclass[a4]{article}
\usepackage[a4paper]{geometry}
\usepackage{cite}
\usepackage[pdftex]{graphicx}
\usepackage[cmex10]{amsmath}

\usepackage{times}
\usepackage{array}
\usepackage{graphicx}
\usepackage{mathtools}
\usepackage{amsmath}
\usepackage{enumitem}

\usepackage{alltt}
\usepackage{shadethm}
\newshadetheorem{comment}{Comment}

\def\boxx{{\vcenter{\vbox{\hrule height.3pt
          \hbox{\vrule width.3pt height6pt
          \kern6pt\vrule width.3pt}\hrule height.3pt}}\;}}

\def\impos{{\;\vcenter{\hbox{\rule{5mm}{0.2mm}}} \vcenter{\hbox{\rule{1.5mm}{1.5mm}}} \;}}

\def\lrarrow{\leftrightarrow \kern-8pt \rightarrow}

\def\2{\frac{1}{2}}

\def\beq{\begin{eqnarray}}
\def\eeq{\end{eqnarray}}
\def\2{\frac{1}{2}}

\newtheorem{definition}{Definition}

\def\lrarrow{\leftrightarrow \kern-8pt \rightarrow}

\def\frightarrow{\rightarrow \kern-11pt /~~}
\def\reducesto{\simeq \kern -3pt >}
\def\intersection{\cap}

\begin{document}
\newcommand{\strust}[1]{\stackrel{\tau:#1}{\longrightarrow}}
\newcommand{\trust}[1]{\stackrel{#1}{{\rm\bf ~Trusts~}}}
\newcommand{\promise}[1]{\xrightarrow{#1}}
\newcommand{\revpromise}[1]{\xleftarrow{#1} }
\newcommand{\assoc}[1]{{\xrightharpoondown{#1}} }
\newcommand{\imposition}[1]{\stackrel{#1}{\impos}}
\newcommand{\scopepromise}[2]{\xrightarrow[#2]{#1}}
\newcommand{\handshake}[1]{\xleftrightarrow{#1} \kern-8pt \xrightarrow{} }
\newcommand{\cpromise}[1]{\stackrel{#1}{\frightarrow}}
\newcommand{\policy}{\stackrel{P}{\equiv}}
\newcommand{\field}[1]{\mathbf{#1}}
\newcommand{\bundle}[1]{\stackrel{#1}{\Longrightarrow}}

\title{{\small Spacetimes with semantics II
    (supplement)\footnote{These notes are a continuation my series
      on semantic spacetimes. This document is inspired by the studies of
      coarse-grained universal scaling in
      cities\cite{bettencourt1,bettencourt2,bettencourt3}, and a
      comparison with models developed over the past decade or two on
      information systems, e.g. \cite{burgessbook2,burgesstheory}.  As
      IT systems grow in scale, is natural to expect a bridge between
      the behaviours of cities, software networks, and other
      functionally `smart' spaces, and one hopes for a better
      understanding of pervasive information technology in social
      contexts. Work on this, from the low level viewpoint, has
      already begin in \cite{spacetime1,spacetime2}. I want to show
      how these views relate.  }}\\On the scaling of functional
  spaces,\\from smart cities to cloud computing}

\author{Mark Burgess}

\maketitle

\begin{abstract}
  The study of spacetime, and its role in understanding functional
  systems has received little attention in information science.
  Recent work, on the origin of universal scaling in cities and
  biological systems, provides an intriguing insight into the
  functional use of space, and its measurable effects.  Cities are
  large information systems, with many similarities to other
  technological infrastructures, so the results shed new light
  indirectly on the scaling the expected behaviour of smart pervasive
  infrastructures and the communities that make use of them.  

  Using promise theory, I derive and extend the scaling laws for
  cities to expose what may be extrapolated to technological systems.
  From the promise model, I propose an explanation for some anomalous
  exponents in the original work, and discuss what changes may be
  expected due to technological advancement.
\end{abstract}

\tableofcontents

%%%%%%%%%%%%%%%%%%%%%%%%%%%%%%%%%%%%%%%%%%%%%%%%%%
\section{Introduction} 
%%%%%%%%%%%%%%%%%%%%%%%%%%%%%%%%%%%%%%%%%%%%%%%%%%

Two recent works have shed light on the role of infrastructure in the
scaling of functional systems.  The observation of universal, scale
free behaviours, both in observed data, and in a `mean field'
description of cities\cite{bettencourt1,bettencourt2}, and the observation, in
biology, of metabolic scaling laws for organisms, following a
century of observation\cite{gwest1}. Both of these were explained from
the behaviours of their internal networks.

Cities are an example of collaborative community networks, where
humans and technology mix within a semantically (i.e.  functionally)
rich space, equipped with infrastructure. One may ask how cities
differ from apparently similar communities across different scales,
including tribes, collectives, companies, software installations, and
even countries.  Understanding the dominant processes that make these
shared environments smart, creative, and productive, is a worthy
investment, given how 21st century life relies on them so much for its
success and survival.

In this note, I review and build on Bettencourt's model of
cities\cite{bettencourt1}, and discuss its implications for pervasive
information technology (IT). By building on the lessons of cities, I
hope to to foster a better understanding of a broader range of
functional systems, especially in information technology, while trying
to preserve the simplicity of Bettencourt's arguments. I begin by
summarizing an interpretation that lays the foundations for a more
microscopic formulation of the model, using promise
theory\cite{promisebook}. The latter may be used to relate outcomes to
intentions and mechanisms, in a way that respects the idea of scaling.

%%%%%%%%%%%%%%%%%%%%%%%%%%%%%%%%%%%%%%%%%%%%%%%%%%
\section{Bettencourt's model of scaling in cities summarized}
%%%%%%%%%%%%%%%%%%%%%%%%%%%%%%%%%%%%%%%%%%%%%%%%%%

This section is a review (and trivial generalization) of the city
scaling arguments, and data used by Bettencourt and collaborators at
the Santa Fe Institute (SFI) \cite{bettencourt1, bettencourt2,
  bettencourt3}, in a form suitable for comparison with other work. In
section \ref{pt}, I propose a deeper justification for the model, with
some embellishments.  Some interpretations may be my own.

\subsection{Scaling phenomenology}

Measurable attributes, of finite dynamical systems, typically scale
in proportion to some measure of their size. Size may refer to the
number of agents $N$ (persons) in the system, or per unit volume $V$
of the system, and $N$ and $V$ may or may not be related. This is a
point of view that is a basic tool of analysis in physical systems.
For cities, $N$ is used as the scaling variable.  Across an ensemble
of many systems of different size, the measurements one obtains may
scale in three broad ways (see figure \ref{scalingz}):
\begin{figure}[ht]
\begin{center}
\includegraphics[width=13cm]{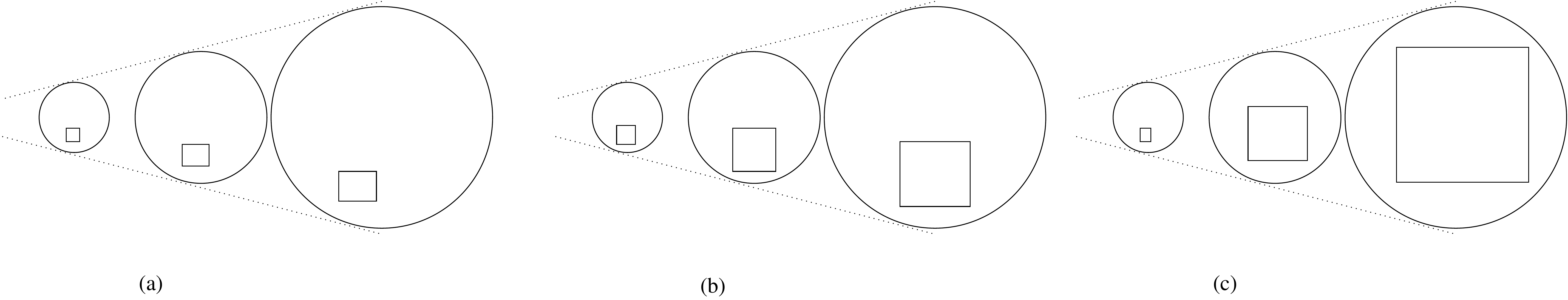}
\caption{\small Scaling of a square quantity, relative to the circular
  system size (a) sublinear, (b) linear, (c) superlinear. If an cost
  is superlinear, to applies a braking force on city
  size.\label{scalingz}}
\end{center}
\end{figure}

\begin{enumerate}
\item Sublinear scaling of quantities $q$ of the infrastructure
  machinery $q \propto N^{\beta<1}$. This indicates economies of
  scale, because, as the system size grows, the cost becomes
  relatively cheaper. In cities, it is observed to apply to the transport
  networks that animate the system (arterial systems, roads, cables,
  petrol stations, etc).

\item Linear scaling, simply proportional to the number $q \propto
  N^{\beta = 1}$.  In cities, this seems to apply to direct
  consumption of goods and resources per capita (jobs, houses, water
  consumption, etc).

\item Superlinear yields of produce or `output', where $q \propto
  N^{\beta > 1}$, which is driven by interactivity between the parts
  of the system and its consequences (wages, rents, patents, crime,
  disease, GDP).  If a process rate is superlinear, then the
  corresponding time for the process to run will be sublinear, and
  vice versa.
\end{enumerate}
It is of particular interest when these patterns seem to apply across
such a broad range of scales. This suggests some emergent
{\em universality}, whose origin and mechanism is worth understanding.  

As part of a protracted project to uncover the behaviours of cities or
urban metropolitan districts, Bettencourt has proposed a mean field model to
explain observed scaling behaviour of certain economic measures,
making only elementary assumptions about the processes
within\cite{bettencourt1}.  The model predicts the main features of
the data, by assuming a dynamical universality, but seems to fall short of
describing the full range of observed scaling exponents $\beta$.
Empirical data from \cite{bettencourt2} revealed sublinear, linear,
and superlinear scaling behaviour in the variety of accessible data.
The data came initially from mostly American cities (see table
\ref{innovtab}), and have since been demonstrated in European cities
in \cite{bettencourt4}. All cities, thus far, have a more or less
comparable level of technological development, and thus fit plausibly
into a statistical ensemble.
\begin{table}[ht]
\begin{center}
\begin{tabular}{|c|c|c|}
\hline
\sc Sublinear & Linear & \sc Superlinear\\
\hline
Fuel sales        & Housing &New patents\\
Fuelling stations & Employment &Inventors\\
Length of cables  & Power consumption &R\&D employment\\
Road surface      & Water consumption &Other creative employment\\
                  & &Disease (AIDS)\\
                  & &Crime\\
\hline
\end{tabular}
\end{center}
\caption{Examples of city outputs with sub- and superlinear scaling per capita,  reported in \cite{bettencourt2}.\label{innovtab}}
\end{table}

Cities are not just dynamical systems, they also exhibit complex
semantics, or purposeful, intentional, patterned behaviours.  In the
physics of inanimate systems, the markers of semantics are comprised
of only a few simple labels; charges, force laws, and interaction
graphs.  These are constant over time, and accepted as universal
`physical law'.  However, in human functional systems, the range of
interactions, and their assumed meanings, is much richer, and may
depend of time and context.  Coarse graining, by creating a mean field
model, is a standard physical approach which eliminates the types,
labels, and other semantics of networks, and exposes the universality
of scaling phenomena. However, this simplicity is a trade-off:
semantics are also what sustains the arrangement and composition of
functional processes, at a deeper level, and could lead to actionable
predictions. 

In these notes, I show how semantics provide some additional
structure and constraints on dynamics, and how both short and long
range interactions may be distinguished through functional dependency.
This leads to a possible explanation for the discrepancy between data and
predictions of superlinear scaling exponents in
\cite{bettencourt1,bettencourt2} using a slightly more detailed model
than \cite{bettencourt1}, based on promise theory\cite{promisebook}.

\subsection{Definitions}

Consider a city, in the coarsest approximation, as a bounded
homogenous mass, sustained by external supplies and internally
generated wealth.  This is analogous to a model of gravity and
pressure versus volume, whose equilibrium defines an average city
size. Bettencourt likens the arrangement to a star, which gravitates
together from the benefits of city infrastructure, and expands through the
accretion of new inhabitants.

The city is populated by $N$ inhabitants, which I will call `agents',
which and may include machines, animals, etc., or any proxies for
human intent that lead to economic output. The data on cities, used in
the comparison, are based on human population numbers, so $N$ will
refer to the people, in the first approximation.  The agents are
distributed within a volume $V$, in $D$ dimensions. Cities are more or
less two dimensional ($D=2$), in spite of high rise regions, because
the more or less 1 dimensional networks connecting them lie mainly in
a plane.

\subsection{Outline of the approach}

\begin{itemize}

\item Let the number of agents in the network, or city community, be:
  \beq N = N_I + N_0 \eeq where $N_0$ is a partial dead-weight
  population that is not interacting with the city infrastructure (as
  user or provider)\footnote{The dead weight could be green spaces,
    but more likely slum dwellings such as those that dominate
    Baltimore.  Hot-beds for crime and shadow economy.  This merely
    interferes with city output as far as the world is concerned.},
  and the functional networks generating the yields are agents from
  the $N_I$ population.  Bettencourt does not distinguish between
  $N_I$ and $N_0$, but it is useful to track this distinction
  throughout the scaling argument.

  A dependence on $N$ can mean two different things, in the formulae:
  an average value of a static population, across an ensemble of
  different cities, or a dynamically growing value of $N$ within a
  single city, over the course of its evolution. The interpretation of
  the results is sensitive to this distinction, requiring some care.
  If universality were completely true, they would be the same.

\item A city is a volume $V$ of agents, accreted from a wider region,
  with finite compressibility, by virtue of some average space
  requirements per person, and a pressure a (non-detailed) balance
  condition, which matches the output shared amongst interactions to
  the resources that can be fed to the connected agents.

\item An estimate is made of the minimum fraction of the volume to be
  infrastructure $V_I$ that could connect the city's agents together
  into a virtual mesh. 

\item The city infrastructure is assumed to be `sparsely' utilized, at
  equilibrium, in the sense that the technology on which ensemble
  cities are based can absorb fluctuations in space and time, without
  signifiant contention or expansion\footnote{In Feynman's words,
    there is `plenty of room at the bottom'. Fluctuations in network
    utilization generally exhibit long-tailed
    behaviour\cite{linked,burgessC8}, so we must be far away from the
    point of collapse, in spite of superficial appearances like
    traffic jams.}, else it would choke to a standstill.  This is a
  technical (statistical) property, which is necessary to achieve
  economies of scale, through spacetime multiplexing\footnote{Arrival
    processes like Poisson and L\'evy distributed events have the
    property that a convolution of multiple flows is form invariant.}.
  A city may appear to be busy, even crowded at certain moments, but
  `sparse' means that it could technically get a lot more crowded, on
  average, without collapsing from congestion.

\item Various functional output yields $Y$, of the city may be
  calculated in a form suitable for ensemble comparison, to determine
  their scaling with $N$. Some outputs stem from individual agents,
  and some are from the interactions between them.
\end{itemize}

\subsection{Explicit and implicit assumptions behind a mean field model}

\begin{figure}[ht]
\begin{center}
\includegraphics[width=9cm]{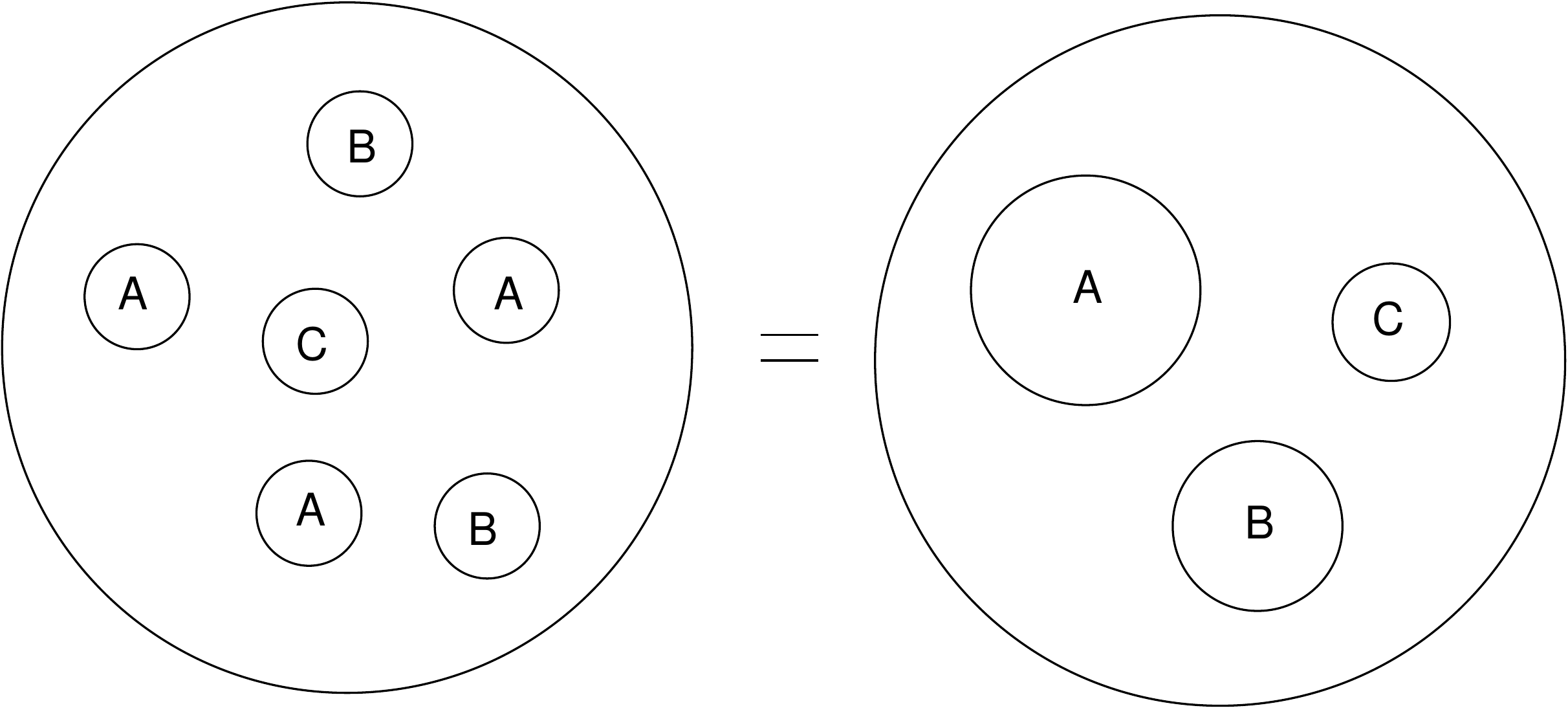}
\caption{\small The yield agency model is a pudding model, in which different
yields may be spread about inside the city bounds, but they may all be lumped
into a single equivalent volume for the purpose of scaling the ingredients.\label{pudding}}
\end{center}
\end{figure}

To make a mean field explanation plausible, over a range of scales and
circumstances, some assumptions are required. The mean field model
should not be confused with a detailed model of physical processes. It
is an effective representation that emphasizes universal aspects of
behaviour.  Expanding on what is stated in \cite{bettencourt1}:
\begin{itemize}

\item The ability to form an ensemble over multiple cities, assumes
  that when similar outputs are promised, they are made by the same
  kinds of agents, with the same basic
  capabilities\cite{Arcaute20140745}.  If one city has a technology
  that makes the same output twice as productive, this will not
  necessarily respect the ensemble's universality assumption, leading
  to anomalies.  The latter assumption suggests that there might be
  difficulties comparing third world cities, primitive societies, or
  fully automated production facilities with more homogenized average
  candidates.

\item The distribution of agents within a city cannot be predicted or
  described by a mean field model. I'll return to this issue in
  section \ref{pt}.  The agents involved in a specific output `yield'
  $Y$ might be concentrated into regions, however, there may also be
  multiple regions with the same role, adding up to the total partial
  volume $V_Y$ used in the expressions. All are lumped into a single
  equivalent volume $V_Y$ for the purpose of total comparison (see
  figure \ref{pudding})..

\item A key assumption is that infrastructure filaments take up only a
  small or `sparse' fraction of volume of the city $V_I \ll V$, but
  can cope with all of its load requirements.  This sparse use of
  resources, with the city volume, will be important to justifying the
  superlinear scaling.

\item In this rendition of the model, I add one trivial extension to
  \cite{bettencourt1}: the infrastructure network connects a fraction
  of $N_I \le N$ of the total population $N$ together, leaving some
  residual number $N_0$ of city dwellers unconnected, or
  non-participatory. This allows us to track the possible impact of a
  non-productive mass.

\item The infrastructure network is not a mesh network itself, but it
  delivers sufficient capacity and interconnectivity to allow a
  virtual mesh of coverage, i.e. any agent can reach any other agent
  with equal average cost.

\item The maximum output of the city community is assumed to follow
  Metcalfe's law, which estimates the productivity of a network
  proportionally to the maximum number of links that can be
  made\cite{metcalfe}. This has been criticized theoretically (for a
  sample see \cite{refmetcalfe1,refmetcalfe2}). However, recently this
  conjecture has received empirical support from social media
  studies\cite{facebooktencent}\footnote{Metcalfe originally assumed
    that value creation would be proportional to $N^2$ while costs
    grew proportional to $N$. The study \cite{facebooktencent}
    indicates that both grow quadratically with network vertex count,
    though these ideas are still
    disputed\cite{refmetcalfe1,refmetcalfe2}.}. I will derive this
  result, and its limitations, from promise theory in section
  \ref{metcalfe}, and also show that this cannot be the only measure
  of productivity to reproduce the data.

\item Let us measure both value and cost in the dimensions of `money'
  $[M]$. In \cite{bettencourt1}, the author uses power (energy per
  unit time) as the proposed currency; however, money might be easier to
grasp for many readers.

\end{itemize}
Regarding the economic output of the city, the model is quite simple.
\begin{itemize}
\item The city's activities yield outputs, labelled by different $Y$.  

\item These come from groups of agents that interact and collaborate
  in an unspecified way.  The group occupies an aggregate volumes $V_Y$, for
  each $Y$ (see figure \ref{pudding}).
These patches are virtually contiguous even when they
are physically distributed, and so they may overlap in space. 

\item By assuming that $V_Y$ is a fraction of the $N_I^2$ interaction mesh, one
  implicitly allows the members of a yield producer to be located
  anywhere within the equal-cost network.

The economic output due to a process $Y$ may be written approximately as:
\beq
Y = g_Y\frac{v^\text{coop}_Y}{V_I} N_I^2,\label{scaling}
\eeq
where $g_Y$ has units of money $[M]$, and $N_I$ is assumed dimensionless.
In many cases we can assume that $N_I\simeq N$, so that the entire city is
active (no dead weight or free riders), but there is no need to make this
identification yet.

\end{itemize}

\subsection{The fraction of volume that is sparse infrastructure network}

\begin{figure}[ht]
\begin{center}
\includegraphics[width=11cm]{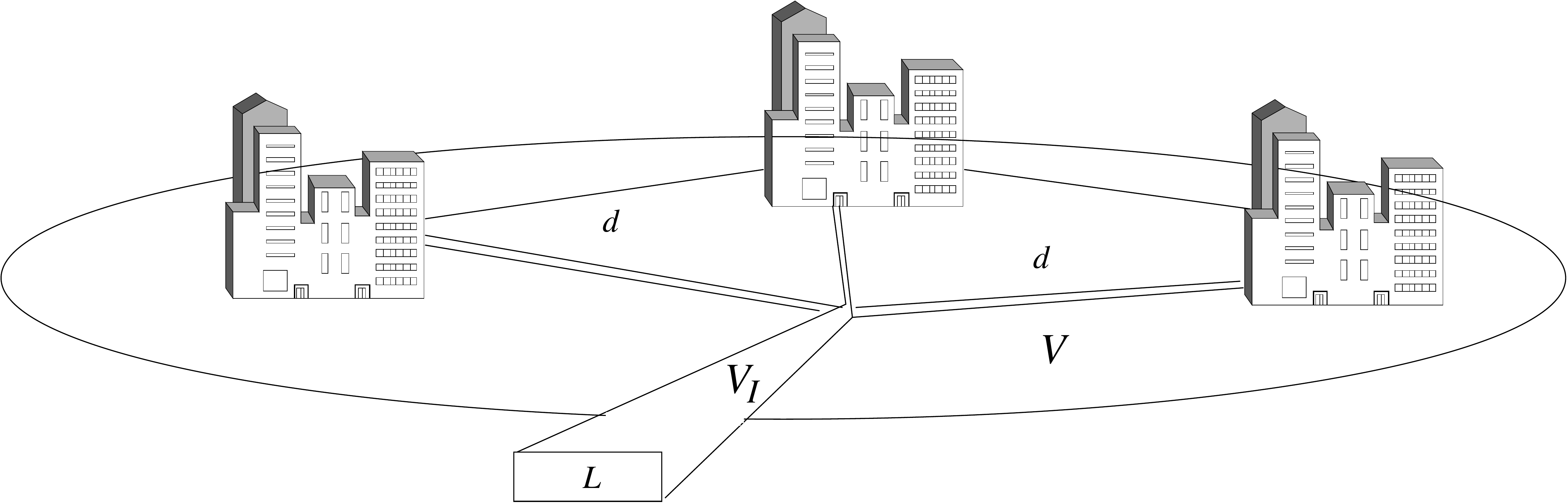}
\caption{\small The volume of the infrastructure sparse network is negligible
compared to the ball of the city.\label{infranet}}
\end{center}
\end{figure}

In the volume of the city, network links are counted using
a continuum approximation in terms of volumes, and fractions of
volumes. This is an unfamiliar step in computer science, but it plays
an important role in deriving the scaling laws, and this
case can offer valuable lessons.  The minimum size of
the infrastructure network can be estimated by squeezing the total
sparse volume into a narrow, approximately one dimensional pipeline,
with a small cross section.  This is only plausible of the network
utilization is really sparse, since then the total interaction can be
compressed into the lower dimensional network, by multiplexing.
The average distance between agents inside the city (in $D$ dimensions) is
\beq
d = \left(\frac{V}{N}\right)^\frac{1}{D}.
\eeq
An infrastructure network has the topology of a graph, in the
mathematical sense, but it may also have sufficient structure to
pervade space\footnote{This was an important argument in deriving the biological
scaling laws\cite{gwest1}.}. It is embedded in a real world volume, and needs to
reach the agents distributed within. If the agents cluster around the
network, the system will remain largely one dimensional; however, if
the network penetrates the space homogeneously (either by wiring or
by the movement of inhabitants who use it), then (again, in the spirit of generality) 
the effect of this `space filling', or fractal 
invasion\footnote{The model cannot formally distinguish between the
  intricacy of the infrastructure itself and the movement of agents
  around it, but it makes sense to assume that it is the motion of
  people and mobile agents that is complex, rather than the system of
  roads and wires of the city. I'm grateful to Luis Bettencourt for
  this comment.}, may be captured by an effective (Hausdorff) dimension $H < D$,
so that we may write the order of magnitude estimate for the
infrastructure volume:
\beq
V_I \ge g_I\left(\frac{V}{N}\right)^\frac{H}{D} L^{D-H} N_I,\label{vie}
\eeq
where $g_I < 1$ is a dimensionless constant that indicates the fraction
of nodes in $N_I$ spanned by the particular infrastructure being considered.
$L$ is some fixed scale with the dimensions of length $[L]$, so that
\beq 
[V] = [L]^D.
\eeq
and $L^D \ll V_I \ll V$.  In other words, the volume is the effective
average linear volume swept out by a fixed cross section $
L^{{D-H}}$, as it feeds into the $N_I$ nodes connected by the
infrastructure\footnote{It is well known that the scaling of ad hoc
  communications networks, where agents are distributed randomly is
  like $\sqrt{N}$. This is easily understood from the spatial
  geometry: mobile phones occupy some approximately two dimensional
  area, so the diameter is of order $N^{\frac{1}{2}}$; alternatively,
  they have average separation $d \simeq V/\sqrt{N}$, so the distance
  across the group is of the order $Nd\simeq \sqrt{N}$.  The linearity
  of the process gets mixed up with the geometry of the embedding
  space.  }.  This has the schematic form of $N_I$ queues that are
serialized paths of dimension $V^{1/D}$.  For $H = 0$ the nodes are
unconnected, for $H=1$ roads are serial or linear, and for $D > H >
1$, the roads or channels have an effective fractal `thickness', from
a coarse-grain perspective (see fig.  \ref{infranet}). It turns out
that we only need to look at $H=1$, as it is serialization rather than
physical dimension that is important.

The assumption that we can squash a volume $V$ into a smaller
linearized volume $V_I$ is the key process, or universal mechanism for
comparison between cities. It expresses what we mean by a `sparsely
utilized' infrastructure, i.e. the gas of inhabitants is somewhat
compressible.  One thinks first of physical channels, such as roads,
cables, and transportation costs etc; however, any serial stream of
work could constitute work process from a number of agents within the
volume $V$. 
Thus we write the infrastructure volume as approximately (rewriting (\ref{vie})):
\beq
V_I \ge g_I V^\frac{H}{D} L^{D-H}\; N_I N^{\frac{-H}{D}},\label{vi}
\eeq
representing $N_I$ serial queues of supporting services, being fed from
a $D$ dimensional region.  The $V^{1/D}$ scaling represents a serialization of the
work from across the homogeneous region.

\subsection{Inflating the community: sustainable city volume at steady state}

Although the city population and volume are presumed to accrete, as
people come, attracted by the promises of the city, a given population
has to be sustainable.  The total population and volume of the city
must satisfy an equation of state, that explains the average balance
of these payments. The scaling of these payments is the same, but in reverse,
so the balance ends up only as a sign to the coefficients.
From the structure, there must be three main
parts:

\begin{itemize}
\item {\bf Agent sustenance}: the existence constraint on agent
  survival, individually, represents a separate concern that does not
  play directly into the scaling of the city (every agent for itself).
  It is represented, implicitly, through the assumption of constant
  $N$.  This has two aspects: the attractive force that brings people
  to the city, and the resources that feed and supply the balance of
  payments. Few cities are self-sufficient within their bounds. These
  aspects remain formally unexplained in \cite{bettencourt1}, and thus
  do not play a detailed role in the data described.

\item {\bf Work output sustenance}: output yields $Y$, produced by the
  various agents of the city, whether individuals or factories, are
  assumed to make the city cash flow positive, together with the input
  of external resources, making that the city is economically viable.
  This feedback is not modelled directly, only assumed by the positive
  signs of the coupling coefficients along links, and the assumption of
  steady state.  Thus, cities might be profitable, or borrowing money
  to reach this steady state. There is no way to capture these factors
  in the mean field approximation.

\item {\bf City volume sustenance}: what we can model is 
the supply of resources must balance
the outputs of the city.
\beq 
\text{Running cost} \le \text{Resource supply in} - \text{Transport cost}.
\eeq 
The running and transport costs are assumed to be low, else the agents
would not be able to sustain their existence. The costs are in the links
and in the linearized supply through the infrastructure.

The feeding of resources (perhaps from outside the city) through the
linearized infrastructure (RHS) powers interactions within the partial
city volumes, over many activities, (LHS), and effectively inflates
the volume of the city by placing a scale $V$ over which resources
must be transported. This yields a simple (non-detailed) balance condition:
\beq
\text{Cost of interaction links} &\le& \text{Supply through infrastructure}\nonumber\\
g_Y\frac{v_Y}{V} N_I^2 ~~&\le&~~ c_Y V^{\frac{H}{D}} N,\label{balance}
\eeq
where we assume that work cost is proportional to the distance
travelled (analogous to $W=F\cdot dl$), and $c_Y$ is the cost per unit
length of path transport along the infrastructure (dimensions $[M]
[L]^{-H}$). The positive coupling $g_Y$ includes the balance of
payments to keep the city viable.  Rearranging this inequality gives
the implied constraint on the sustainable volume:
\beq
V \ge a \left(\frac{N_I^2}{N}\right)^\frac{D}{D+H}\label{vol}
\eeq
where $a = (g_Yv^\text{maintain}_Y/C_Y)^\frac{D}{D+H}$.

\end{itemize}

\subsection{Scaling of spatial infrastructure and output yields}

Substituting the steady state volume of the city into the expression for the
infrastructure volume, the model now predicts the three kinds of scaling from the introduction:
\begin{enumerate}

\item {\bf Sublinear}. The scaling of infrastructure itself
in terms of the volume, we get:
\beq
V_I = g_I a^{H/D}\left( \frac{N_I^2}{N}\right)^{-\frac{2HD+H^2}{D(D+H)}}\label{visub}
\eeq
setting $D=2$ and $H=1$ into (\ref{vi}), 
\beq
V_I  \simeq \left(\frac{N_I^2}{N}\right)^{\frac{5}{6}}.
\eeq
For pervasive $N\simeq N_I$, this yields
\beq
V_I  \simeq N^{\frac{5}{6}},
\eeq
giving the sublinear scaling observed by \cite{bettencourt1}.
When the infrastructure cost is basically absent, this scales like `every man for himself',
like an ideal gas of non-interacting agents.

\item  {\bf Linear}. For individual agent consumption, the scaling is trivially linear, by assumption,
both inputs (consumption $C$) and outputs $O$.
\beq
C_- &=& e_- N\\
C_+ &=& e_+ N
\eeq
where the dimensions $[e] = [M]$ are of money.

\item  {\bf Superlinear}. The positive economic yield of a process in the city, 
due to interactions may be written as a fraction of the possible
$N_I^2$ output that can be channelled through the volume $V_I$, with
a different constant of proportionality for each output:
\beq
Y^+_Y = G_Y \frac{N_I^2}{V_I},\label{yy}
\eeq
where $G_Y$ is assumed positive, absorbing the costs of interaction,
and from (\ref{vi}) we have an expression for the infrastructure volume,
up to invariant unknowns, which are simply constants that do not depend on
the state variables $N$ or $V$. So, substituting (\ref{visub}) into (\ref{yy}),
\beq
Y^+_Y = \left(g_I G_Y  L^{H-D}\right) V^{-\frac{H}{D}} N_I N^{\frac{H}{D}},
\eeq
and substituting for volume in (\ref{vol}), since it also depends on $N_I$:
\beq
Y^+_Y = \left(g_I G_Y L^{H-D} a^{-\frac{HD}{D(D+H)}}\right)  
N^{\frac{2HD+H^2}{D(H+D)}} N_I^{\frac{D^2-HD}{D(D+H)}},\label{yyy}
\eeq
If we substitute $D=2$, and $H=1$, this scales as
\beq
Y^+_Y \simeq N_I^{\frac{7}{6}}\left(1+\frac{N_0}{N_I}\right)^{\frac{5}{6}}.\label{super}
\eeq
And if we further assume that the infrastructure network is pervasive, so that
$N \simeq N_I$, then
\beq
Y^+_Y \simeq N_I^{\frac{7}{6}} \simeq N^{\frac{7}{6}}.
\eeq
This reproduces the superliner scaling identified theoretically in
\cite{bettencourt1}, and this result matches about half of the
superlinear scaling data quite well. Other data show significantly
higher values for the scaling.  If the infrastructure network is
insignificantly small $N_I \ll N_0$, then there would be binomial
corrections to a $1/N$ scaling.  The `dead population' reduces the
power law slightly (in binomial corrections), so we could expect
processes, for which most of the city cannot contribute, to scale
below the $7/6 = 1.16$th power, thus this cannot explain the higher
scaling exponent. An enhanced explanation is proposed in section
\ref{recursion}.

\end{enumerate}

\subsection{Remarks about the calculation}\label{balloon}

In spite of a smattering of city related narrative, this calculation
is based on a very simple and universal argument about resources
exchanged between a $D$ dimensional volume and a one dimensional
supply network. There must be sufficient spacetime volume associated with
ensemble-standard infrastructure to accommodate growth in $N_I$, or an
increase in density of the users, without saturating it.  Multiplexing
in space and time, is they key to this.

\begin{itemize}
\item Measures, relating to the output of self-sufficient agents, scale
linearly, as one would expect. If we double the number of suppliers,
there is twice as much availability. 

\item Shared resources may exhibit economies of scale if they become
  relatively fewer per person, as cities grow, without impairing
  output.  These economies of scale seem to be quite well captured by
  the model, using a value of $\beta=5/6 \simeq 0.8$ to match data,
  which suggests that the result makes sense both for $N$ interpreted
  as an ensemble average and as a growth parameter over time.

\item Finally, there are superlinear outputs, whose behaviour is more
subtle.  A production output $Y$ may be a fraction,
not of $N$, but of $N^2$ because the maximum output can depend on the
interactions between agents, and the agents working alone.  Output may
or may not depend on the volume of the city, i.e.  perhaps only the
number of agents or currents in the pudding, and perhaps the space
they occupy in the course of their interactions; this depends on the
kinematics of the detailed processes.
\end{itemize}
For every scaling benefit, there may also be scaling costs, with the
opposite sign: contention
for shared resources, spiralling costs of equilibration, etc.
Some general comments about the mean field approximation:
\begin{itemize}
\item People have different jobs, capabilities, and habits. In the
  mean field representation, such detail is not represented
  explicitly, but this does not imply that they don't exist.  Indeed,
  they must be present to account for different flavours of output.
  However, this is not a one-to-one association. Formally, the outputs
  are fractions of the total amount of possible produced work,
  averaged over all specializations, mechanisms, and functions. To
  understand why this is plausible, we need to understand that outputs
  are the result of combinations of jobs, in collaboration (see
  section \ref{whypercolate}). They are assumed equally likely on average across
  cities of the ensemble.  The diversity of a city may increase with
  size\footnote{This could be checked by comparing the size of yellow
    pages directory for a community to its white pages directory.},
  but this does not matter as long as we assume that all jobs are the
  same.  The data suggest that this does matter (see section
  \ref{recursion}).

\item A city must have `potential barriers', or entry costs for
  construction and community building, that depend on the level of
  technology available. This is also not represented in the model.
  There might be significant debt associated with building, which is
  invisible in this picture. These are transient responses, invisible
  in steady state behaviour.

\item The technology at the time of building ought to affect the
  density of the city too: as transport improves and gets cheaper, it
  enables greater distances to be covered in the same amount of time,
  i.e. lower density. Conversely, it enables more efficient transport
  of provisions to sustain a high population density. One might not
  expect old cities like Rome or London to be immediately comparable
  with Brasilia or Shenzhen\footnote{Just 30 years old, `Shenzhen
    speed' is the stuff of legend in China.}.  Similar questions could
  apply to the pace of life. Does this depend only on the size of the
  city, or also on cultural norms?  What if we compared Dakar with
  Seoul, for example? The scaling depends only on relative density
  however.

\item To be comparable in an ensemble average, one could expect that
  cities would need to be comparable in productive technology,
  composition and wastage. A machine could do many times the work of
  a manual laborer, for instance, so it would not be fair to compare a
  city with mechanization to a city of horses and carts. The data in
  \cite{bettencourt2} were taken from mainly American cities, which
  have a relatively uniform level of technological infrastructure, and
  arise from similar epochs.  There is no particular reason why they
  would be comparable to the slums of Mumbai. But this remains to be
  discovered.

\item By distinguishing between $N$ and $N_I$, it is possible to see
  how about the productive capacity of the city could be altered by
  the presence of unproductive regions. From the expansion of
  (\ref{super}), it would seem that the higher the proportion of the
  population that cannot contribute to a process, the lower the
  scaling power of outputs might become.

  It is known, for instance, that a typical `80/20 rule' (power law
  behaviour) is almost universal for networks, i.e. 80 percent of the
  yield typically comes from 20 percent of the
  agents\cite{albert1,linked}, and the effect of certain hubs on
  transmission is crucial to understanding effective partitioning of
  regions\cite{grapharticle}.  This suggests that a graph theoretic
  explanation can add to the simple mean field picture.
\end{itemize}
Coarse graining extracts only the most generic features of a system,
in the limit of large $N$, i.e. at the coarsest scales.  It is unable
to probe the separation of such scales, by weak local interactions, or
distinguish long and short range interactions.  At smaller range, the
semantics of interactions often become vital to the functioning of a
system. In physics, `semantics' are simple and mostly fixed by
`physical law', e.g.  they manifest as different `charges', `forces'
or allowed interaction types. In a human-technology system, however,
there are many more flavours of interactive behaviour that may be
distinguished, and their number and patterns might even change over
time.

The urban scaling predictions reviewed here have been partially
matched to data, with $N$ on the order of $10^3$ to
$10^7$\cite{bettencourt1}; this offers strong evidence that underlying
semantics of comparable cities are unimportant at these scales.
However, the argument above underestimates some of the data
significantly, especially where the data exhibit a high level of
uncertainty, estimated in the margins for $\beta$. This suggests that there is
something not captured adequately by the preceding argument.

There are two possibilities: either inhomogeneities across the
ensembles are significant, or the outputs do not all follow a single
model, and we are looking for either an addition or a refinement.  I
shall try to shed some light on these issues, in the next section, by
deriving the model from a more microscopic model, using promise
theory.

\subsection{Does innovation characterize superlinearity?}\label{creative}

Some aspects of the data, analyzed in \cite{bettencourt2}, are
suggestive of a fit with the single universality model in
\cite{bettencourt1}, but not all.  The authors attributed the
superlinear scaling quantities to creativity or
innovation\cite{bettencourt1,bettencourt2}, though the link between
this explanation and the calculation of the exponent is incomplete.
Below are approximate representations of few samples of the superlinear scaling data from
\cite{bettencourt2} (see reference for details of the numbers):
\begin{center}
\begin{tabular}{c|c|c}
\sc Measure & \sc Approx. Average $\beta$ & \sc Source\\
\hline
Wages & $1.12 \pm 0.02$ & USA \\
GDP  & $1.13-1.26 \pm 0.1$ & EU, Germany, China\\
Patents & $1.27 \pm 0.02$ &USA\\
Private R\&D employment & $1.34\pm 0.05$ &USA\\
R\&D employment & $1.26\pm 0.1$ &china\\
R\&D establishments & $1.19 \pm 0.03$ &USA\\
AIDS cases & $1.23\pm 0.05$ & USA
\end{tabular}
\end{center}
As pointed out by the authors, many of the numbers have a high level
of uncertainty, due to the difficulty of fitting data from disparate
sources\footnote{My hat goes off to the authors for making this effort though!}.  Even with
generous margins, the single predicted value of $\beta = 7/6 = 1.16$
does not plausibly agree with all of these measures, however, and the
deviation seems to not be attributable to a normal variation.  The
data in the table below, by contrast, were collected in an independent
study in the UK that attempted to verify the superlinearity hypothesis
in only a single measure: patents (see reference
\cite{Arcaute20140745} for details):
\begin{center}
\begin{tabular}{c|c|c}
\sc Measure & \sc Approx. Average $\beta$ & \sc Source\\
\hline
Patents & $1.13 \pm 0.1$ &UK all cities\\
Patents & $0.99 \pm 0.1$ &UK small cities\\
\end{tabular}
\end{center}
The authors of this study point out the difficulty in defining what
constitutes a valid city for the purpose of ensemble comparison.
Those values are significantly lower than those in the first (SFI)
table.  While the SFI data for patents do not agree with Bettencourt's
model, the first set of UK data agree much better (1.13 is close to
1.16), but the second restricted set do not (0.99 instead of 1.16).
These discrepancies warrant an explanation.

From the perspective of promise theory, which I will apply in more
detail in the next section, one thing stands out about the measures in
the first table: the outputs are of two qualitatively different
types: some measures are `promises' and some are `agents'.
\begin{itemize}
\item Patents, wages, GDP, disease, are related to the promises of
  an output, i.e. produce that relate to a production process.

\item Jobs related to research are not outputs but occupations.  They
  represent the state of agents. They are not (directly) the result of
  a process\footnote{They could conceivably be related to a training
    process, but then should be be counted as population or output?}.
\end{itemize}
If we consider the version model of \cite{bettencourt1}, there are two
freedoms that remain to alter (\ref{yyy}). One is to imagine the existence
of a large fractional $N_0$ population (e.g. by invoking the 80/20
rule for the distribution of outputs), relative to the population
$N_I$, involved in making each particular promise.  However, this
would have the effect of reducing the value, so, while it might
potentially explain the UK data, it could not explain the SFI
data. The second is to assume $H>1$, which suggests a
significant self-similarity in the infrastructure
that relates to patent production. However, this does not seem credible
as the infrastructure that enables patents is principally researchers,
which do not think in paths that fill space. Something about the
universality assumptions need to be reconsidered. A possible resolution
is provided in section \ref{recursion}.

One clue to these behaviours could be that the selections represent
highly specialized sub-populations, rather than homogeneous fractions.
Recall that the way superlinearity arises is that an efficiency of
scale effectively gets better at large $N$, leading to an effective
amplification with the size of the population.  Wages and GDP to
involve the largest fraction of the population of a city community.
All the other measures are based on highly specialized populations to
which few contribute.  Another point is that patents are not merely
the result of an economic process, but are sometimes weapons used politically
and strategically as part of non-collaborative economic warfare
between companies.  This suggests that semantics
would play a role in their explanation, and potentially skew some data
(cities with companies like Apple and Samsung, well known for
software patent fights\footnote{Software patents usually have low
  production costs, as they are often trivial and frivolous
  inventions.} might appear differently).

In section \ref{recursion}, a generic possible explanation is
proposed, based on the generalized semantics of scaling agents into
clusters (superagents). If one distinguishes agents from their promises, then
staged efficiencies, arising from functional
dependency, can be compounded or reduced, altering the $\beta$ value
hierarchically.  \beq \text{Infrastructure} \promise{\times 1.16}
\text{Superagents} \promise{\times 1.16} \text{Produce/Output} \eeq To
derive this plausibly, we need to formulate a simple promise theory of
communities.

\subsection{Gunther's `Universal Scaling Law' and spacetime involvement in scaling}\label{uslsec}

Before looking at a deeper model, I want to comment on another scaling
model, from the world of information technology (IT).  Superlinear
scaling has also been observed in high performance cluster
computing\cite{gunther3}, albeit for a different reason. Most IT models are
essentially one dimensional in nature, describing serial processes, divided into
parallel one-dimensional threads. The amplification of output in this kind of flow
network is based on queueing results, which may approximated
quite well by a simple current flow models, when workloads patterns
are sparse and evenly insensitive to mixing (e.g. Poisson, or L\'evy
processes)\cite{burgesslongversion}.

\begin{figure}[ht]
\begin{center}
\includegraphics[width=14cm]{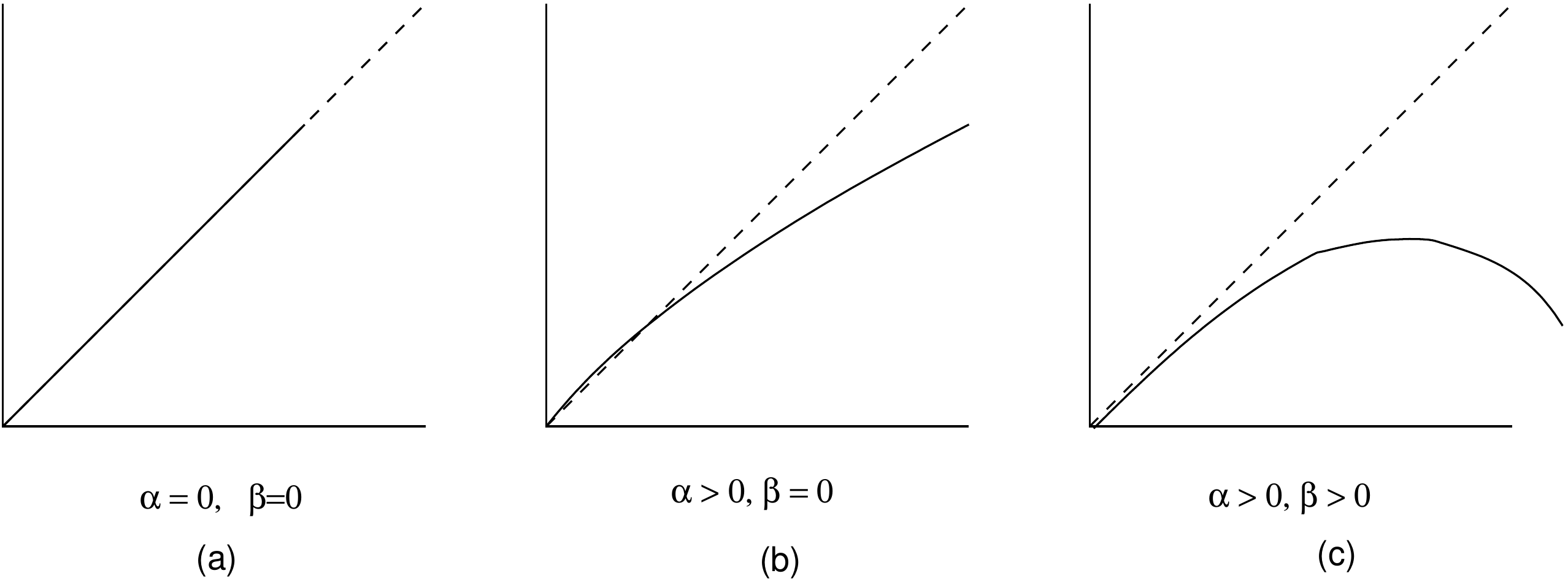}
\caption{The Gunther universal scaling law for its control parameters.
(a) linear scaling, (b) cost of sharing resources and diminishing returns from contention,
(c) Negative returns from incoherency and the cost of equilibration.
\label{gunther}}
\end{center}
\end{figure}

Amdahl's law, and Gunther's generalized `universal' scaling
law\cite{gunther1,gunther2,gunther3}, describe the `speedup', 
or amplification of output, in a basically serial process, by the addition of 
workers. Jobs may be speeded up, but are ultimately limited by the serial
coordination of work. Suppose we have $N_L$ local, parallel worker threads; then this takes the 
general parametric form:
\beq
S(N_L) = \frac{N_L}{1 + \alpha(N_L-1) + \beta N_L(N_L-1)}\label{usl}
\eeq
The output rate for a collective of $N_L$ agents is $S(N_L)$ times that
of a single agent (see appendix \ref{notscin}). The two terms in the denominator represent two kinds 
of network process, at close range. The linear term is a bottleneck term, where
all the agents contend over a limited shared resource. As long as $\alpha > 0$,
scaling will be sublinear. The quadratic term is the cost of making the mesh of
agents agree about something at each stage of the process. This is a cooperation, or
`homogenization of information' (coherency) term, which puts a sharp brake on scaling.
For example, in database or cache replication, information has to be replicated 
consistently, which is expensive.  This term reduces the speed
up to the point where performance can actually grow worse with increasing $N$ (see figure \ref{gunther}).

The expression (\ref{usl}) cannot exhibit a superlinear result for
positive coefficients. Nonetheless, superlinear scaling has been observed
in computing clusters\cite{gunther3}. How can these facts be resolved?
Gunther argues that superlinear scaling is not possible without
violating work conservation; but, by artificially making $\alpha <
0$, as a parametric fit, it is possible to simulate higher dimensional
effects in this one dimensional projection.
This is a one dimensional view of a process that is actually two or
three dimensional.  Datacentre clusters, with dense interconnect
networks approach mesh-levels that fill the effective volume of the
datacentre during their internal processes\footnote{This is called
  East-West traffic in the IT industry.}. As this trend continues, 
we can expect datacentres to behave a lot like the model of cities discussed.

The appearance of superlinear scaling came as a shock in IT, because
there is no way to understand $\alpha < 0$ from a microscopic model.
However, it can be understood as a renormalization effect, i.e. an
effective parametric representation of the projected output.
Qualitatively, the serial limit on scaling could be cheated as
follows, in a sparsely utilized system.  
When the average state of the system is highly underutilized
to begin with, then some of the efficency can be regained by close
packing of the utilization\footnote{This is the motivation for
  packetization (atomization) of networks, and context switching in time sharing
  operating systems.}.  If dense packing can absorb the increased
size of a system (like a city or computational process), then that
economy of scale can enable greater output for a relatively smaller
infrastructure (renormalized negative growth). Could this lead
to a superlinear speedup? The phenomenon of packing is related to 
queue parallelization, where for example $G/G/N$ is provably
more efficient than $G/G/1 \times
N$\cite{kleinrock1,burgessbook2}. This is because wasted
idle time can be eliminated by arranging by an efficient packing of work.  
This cannot persist for ever, as eventually the sparse utilization of the
limiting work agents must
fill up to capacity. As it does so, the amount of contention
($\alpha$ for shared bottlenecks, and $\beta$ for equilibration of state\cite{siri1})
must increase rapidly. The response time $R$ of a queue (proportional to its
average length divided by the dispatch rate $\mu$ takes the form:
\beq
R \simeq \frac{E(n)}{\mu} = \frac{1}{\mu-\lambda}
\eeq
This queue is only stable when $\lambda \ll \mu$, indicating sparse
usage. The scaling indicated in the city results indicates cities that
have not peaked in their infrastructure utilization yet. 
Superlinear scaling sounds like a good thing but it is unstable, as
the queueing projection illustrates.

The rational queueing expression, in Gunther's Universal Scaling Law
(\ref{usl}), can never explain fractional scaling exponents seen in
cities, but it can demonstrate some projected superlinearity with $\alpha <
0$.  To feed superlinearity, we need something more than parallel
serial processes where the work is done by $N$ point sources.  
Only if the work is done by $N^2$ interactions can a partial efficiency
make the exponent greater than unity.
To get this, we need to feed higher dimensional volumes into
lower dimensional volumes. This is what is going on in the mean field
theory of \cite{bettencourt1}.
\beq
\text{Interaction output}  = \text{Maximum output}(N^2) \times \text{Fraction of infrastructure involved}(N)
\eeq
From a graph theoretical perspective, this is a change in average connectivity of the
infrastructure network (i.e. the average degree of nodes $k$\cite{burgessbook2}).
If the fraction is a fraction of a volume rather than a line or a number, there
are dimensional exponents involved, which represent the contact efficiency
by close packing the city population. With more dimensions, a larger surface area
can be used for interaction. If we assume that the infrastructure network is
sufficiently dense that it reaches almost everyone, then this continuum
approximation is reasonable.
\beq
\text{density of infrastructure users} = \frac{N_I}{V_I} = n_I\;N^{\delta(D)}.
\eeq
where $\delta(D) = 1/D(D+1)$, for $H=1$.
Recalling that this volume is really a continuum approximation of a network
of nodes, this translates into an average node degree utilization (or locally used connectivity) 
within the infrastructure channel\footnote{Note that this is not a real connectivity, which
has to do with the number of nodes, but a kind of close-packing of the sparse interactions
that occur between the nodes into the infrastructure stream.}
\beq
k(N) = \frac{N_I}{V_I} = n_I\;N^{\delta(D)}.
\eeq
Assuming the infrastructure is pervasive so $N\simeq N_I$, the equivalent
serialized infrastructure volume, for a single
process, is something like:
\beq
V_I &=& \left(\frac{V}{N}\right)^{\delta} \times N_I \simeq N^{1-\delta} \times \text{cross section},\\
&=& \text{capture volume per agent} \times \text{span of agents} \times \text{cross section},\nonumber
\eeq
Using this volume, instead of the total volume of the system (city,
community, etc), recognizes two things.  First that cost of the
infrastructure is much less than that of the entire system; and,
second, that it is the serialization of the sparse resource over a
standard cross section that we want to use for comparable work output.
This is like fitting the sparse output volume of the city into an
idealized serial stream of fixed width to see how its length scales
with the number of inhabitants\footnote{A simple analogy is to think
  of a tube of toothpaste. The toothpaste comes out in a one
  dimensional stream of fixed width, but we are forcing the output of
  a three dimensional tube through this portal, and asking: how does
  the amount that comes out increase with the size of the tube if we
  squeeze it in the same way? By fixing the cross section, we can
  compare different tubes, or different cities.}.
The efficiency comes from being able to use more of an unexpected fixed cost,
sparsely utilized resource along with other economies of scale. The net result is
an amplification of the output by $\delta$:
\beq
\text{Interaction related output}\; Y = \frac{\text{const}}{V_I} N_I^2 \simeq Y_0 N^{1+\delta(D)}.
\eeq
The numerator is unexpectedly constant, but the infrastructure volume
scales sub-linearly, the net output appears superlinear, with these
assumptions.  The question is how do we know if these are the same
assumptions as the used for the measurements?

Modern datacentres and networks at scale have multiple redundant paths
that make their interconnection networks space filling (e.g. Clos
structures\cite{certainty}).  This brings higher dimensional scaling
issues into the picture.  The general principle is one of close
packing of utilization.  When dependencies scale more favorably than
the contended processes that rely on them (relatively speaking), each
process gets a larger share of the shared resource, and is accelerated
for a while, provided the total utilization remains low.

%%%%%%%%%%%%%%%%%%%%%%%%%%%%%%%%%%%%%%%%%%%%%%%%%%
\section{A promise theory derivation of the city model, and beyond}\label{pt}
%%%%%%%%%%%%%%%%%%%%%%%%%%%%%%%%%%%%%%%%%%%%%%%%%%

I now want to show that we can re-derive the scaling, and indeed the
model of \cite{bettencourt1}, from using a quasi-atomic theory of a
network, taking into account some of the more important semantics to
see how universality emerges, and where its limits might lie.  Promise
theory strikes a balance between semantics and dynamics, and thus
between coarse graining and the chemistry of different functional agents.

\subsection{Brief recap' of promise theory}

Promise theory is a formalism that describes dynamics of atomic, black-box
agents, alongside their broad functional semantics, i.e.  with their
intentional behaviour. The default or ground state of agents is one in
which they make no promises, and are independent or `autonomous', i.e. each agent is
self-sufficient and controls its own internal resources, and each
agent can make promises only about its own capabilities and
intentions\footnote{From a physics perspective, promises look a lot
  like a rich array of charge flavours, for exotic forces, and the
  network looks a lot like a force field.}. Assuming that an agent is
not deliberately deceiving, a promise may be considered the best available
local information about the likelihood of an outcome.

Promise theory can be understood in a number of ways. For the present
purpose, it can be thought of as a labelled graph theory, with some
rules and constraints on interpretation.  A promise from an agent
$S$ to one of more agents $R$ takes the form
\beq
S \promise{b} R
\eeq
where $b$ is the body of the promise. The promise body, explains what
is being promised, and has polarity (+ to offer, and - to accept), as
well as type $\tau(b)$ and constraint $\chi(b)$.  All agents are
considered to be indistinguishable to begin with, and these agents may
work together to form aggregate `superagents', by promising
cooperation\cite{promisebook}.

\begin{figure}[ht]
\begin{center}
\includegraphics[width=5cm]{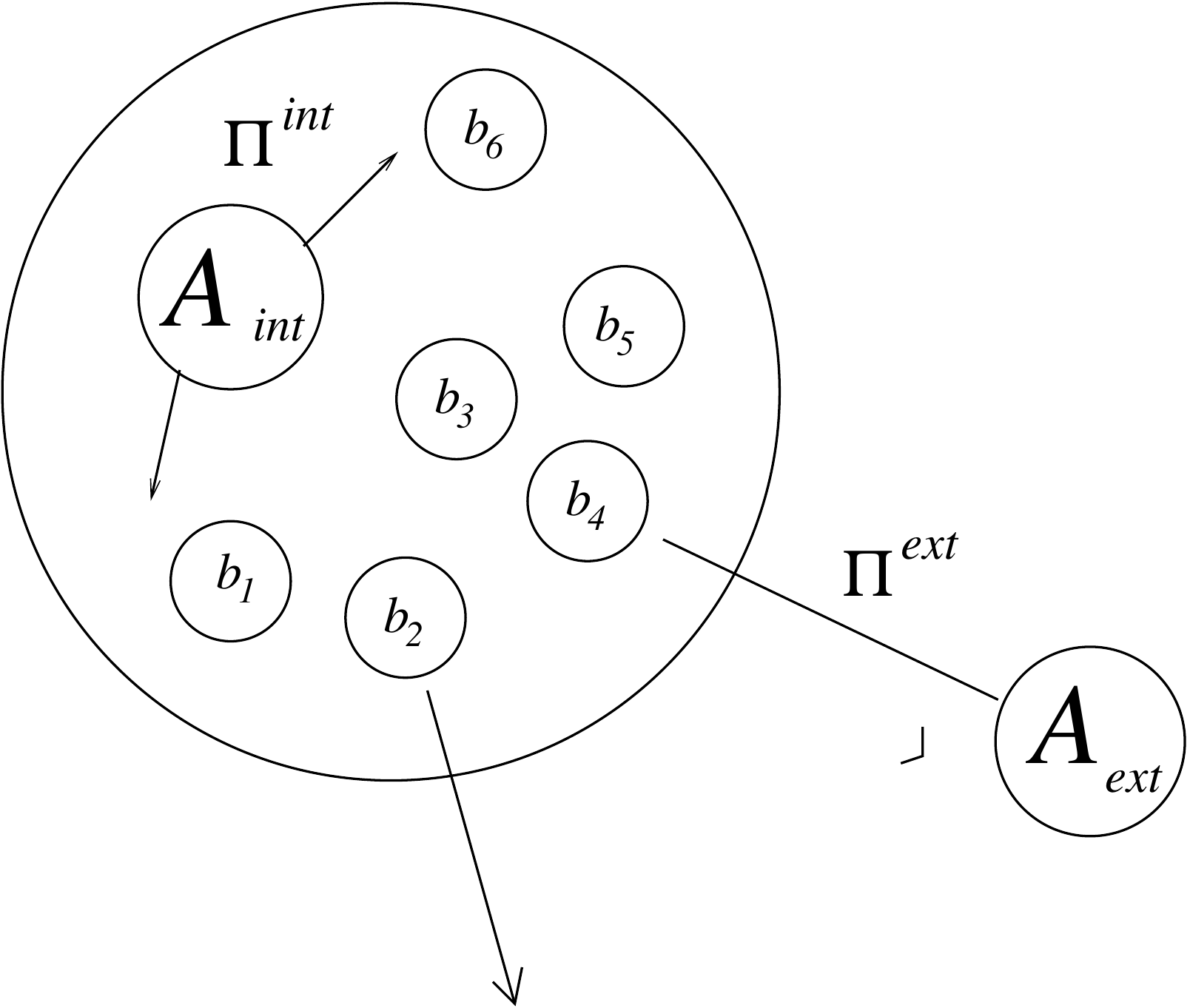}
\caption{\small What is an agent? Agents aggregate to make superagents, with new promises
possible at each scale of agency. So what we consider to be an agent, depends on
the scale at the networks under consideration.\label{coop}}
\end{center}
\end{figure}

In a community or city, we might imagine that the atomic agents are
persons, machines, and even software, acting as proxies for human
intentions.  An aggregate `superagent', such as a company can
collaborate to keep new promises (see figure \ref{coop}). Each
aggregation into a superagent enables new promises to exist that
cannot be attributed to any of its components.  To understand output
yields, the key is to ask: what kind of agent is responsible for an
output? Is it a single point source agent, or a distributed
superagent, with an internal collaboration mesh? In either case,
collaboration has a finite range.
In summary:
\begin{enumerate}
\item Agent types are distinguished by the promises they make. 

\item The way that agents make use of one anothers' intent through
  promises is what we mean by agent semantics.

\item The outcome of a promise is deduced by observation; this is called
an assessment in promise theory.
An assessment by agent $i$ about a promise $\pi$ is written $\alpha_i(\pi)$.

\item  A link between two agents requires a promise by both parties:
one to make a service offer (+), and the other to accept it (-). 
\beq
\left.\begin{array}{c}
S \promise{+b} R\\
R \promise{-b} S
\end{array}
\right\}  ~~~= \text{unidirectional transfer}
\eeq
This kind of binding is the basis for determining the coupling strength $g_Y$
for the output yields.
\end{enumerate}
Promise theory is a theory of incomplete information, and embodies controlled
coarse graining over semantic scales.

\subsection{Agents and super-agents, long and short range interactions}

Consider a set of agents $A_i$, where $i$ runs over the population of
the system (city, datacentre, etc), and all machines and proxies for
human intent.  In order to collaborate, agents need to make some basic
promises\cite{promisebook}.  Every promise may be either kept or not
kept, and the average value needs to be self sustaining.  Each
autonomous agent thus has a balance of payments to consider. It needs
to accept fuel, food, energy, money, etc.  Dependencies also include
raw materials which have to come from outside. If agents can cache
resources they can maintain weak coupling, else they are strongly
dependent on their environments. This applies to energy, supplies, and
also inputs of information and ideas.
Promises made directly between agents are called short range.
\begin{definition}[Short range interaction]
A binding between adjacent agents $S$ and $R$ of the form
\beq
S &\promise{+\tau,\chi_+}& R\\
R &\promise{-\tau,\chi_-}& S
\eeq
where $\tau$ is the same in both promises, and $\chi_+\intersection\chi_- \not=0$.
\end{definition}
A promise may be long range if it is non-local, i.e. it couples several
agents together, or employs intermediaries.
\begin{definition}[Long range interaction]
A binding between adjacent agents $S$ and $R$, through an intermediate agent $A$
\beq
S &\promise{+\tau,\chi_+|d}& R\\
R &\promise{-\tau,\chi_-}& S\\
I &\promise{+d} S&\\
S &\promise{-d} I&
\eeq
where $\tau$ is the same in both promises, and $\chi_+\intersection\chi_- \not=0$.
\end{definition}
Note that there is no a priori notion of distance in a graph, other than the
number of interactions or hops between agents nodes. The familiar notion of distance comes about
from embedding a graph in metric space, which in turn is related to a continuum
approximation.

At any scale, a promising agency that plays a functional systemic role
makes promises of the following general forms.
\beq
A_i \promise{-f} A_{\rm ext}, ~~~\forall ~ i.
\eeq
There must be an external agency acting as a source of this fuel $f$, providing
\beq
A_{\rm ext} \promise{+f} A_i, ~~~\forall ~ i.
\eeq
Any agent $A_i$ may depend on pre-requisite promise of dependency $D$, provided by another, 
in order to provide service $+S$; according
to the assisted promise law\cite{promisebook}:
\beq
A_i \promise{+S|d} A_j, A_i \promise{-d} A_k \simeq A_i \promise{+S} A_j
\eeq
provided
\beq
A_k \promise{+d} A_i.
\eeq
In this way, agents have probes, services, skills (+), and needs or
receptors (-) that can unlock or catalyze their functionality.  The
expresses its exterior promises outwardly, e.g. a door handle's
function is recognized by its shape, just as a car and its promises
are recognizable by its exterior structure. Interior promises might be
involved in making the exterior ones, but these are not generally
visible at super-agent scale.  The basics of scaling semantics and
agency are laid out in \cite{spacetime2}.

\subsection{Promise networks: functional interactions}

Functional networks have two aspects to their productivity, which can be described by:
\begin{itemize}
\item Replication of agent output: dynamics, economics.
\item Combination of used services, ideation by mixing: intent, semantics, fitness for purpose.
\end{itemize}
It is the combination of these that leads to the understanding of fully functional scaling.
Basic communication and supply infrastructure networks enable
interactions between any pair of agencies, but specialist functional
networks are typically small and disconnected\cite{siri3}. They relate
specific promises or services that are constrained by operational
semantics.  The scaling of such networks has been examined in
\cite{spacetime2}.  It has a few aspects:

\begin{itemize}
\item Agents keep promises at scale by individually promising similar
  capabilities in parallel, e.g.  Amdahl's and Gunther's laws of
  scaling\cite{gunther1,gunther2,gunther3}. A promise that is
  considered kept by a single agency scales by having multiple
  sub-agents form a `superagent'. Now Bettencourt's insight reveals
  how output can also scale in a non-parallel
  fashion\cite{bettencourt1}.

\item When one agent depends on a promise being kept by another, in
  order to keep a promise, this creates a serial dependency,
  introducing a queue and a handling rate scale.  The agents must be
  able to understand one another, with common language, else they are
  partitioned. Partitioning cuts off long range interaction, and
  promotes long range diversity, like cultural and specialist
  diversity\footnote{The lack of a common language is effectively a
    channel separation, disconnecting networks into separate branches.
    In communication networks, channel width is sometimes shared
    between different partitioned agencies by using non-overlapping
    frequency ranges.  This is called Frequency/Wavelength Division
    Multiplexing (FDM).  It is a form of multi-tenancy.}.

\end{itemize}

\begin{figure}[ht]
\begin{center}
\includegraphics[width=4cm]{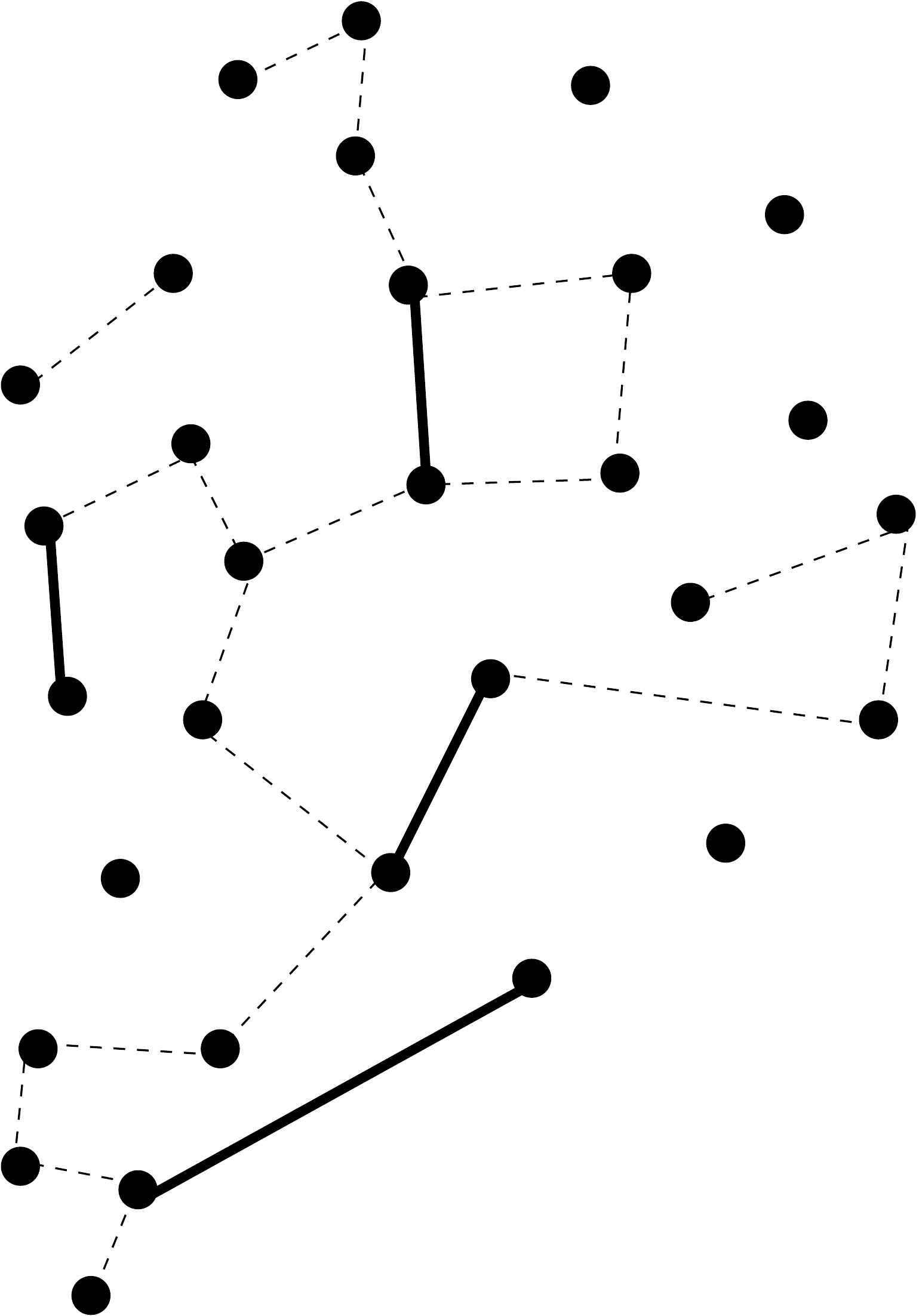}
\caption{\small No single type of promise binding (dark lines) leads to
  percolation of value in the promise graph. However, with conditional
  dependency, and sufficient diversity and homogeneity, there can
  still be effectively close to $N^2$ links whose value converts into
  a common currency.\label{percolate}}
\end{center}
\end{figure}

\subsection{Forces that condense a cluster}

A hypothesis of promise theory is that one may define a notion of
force for agents, which is attractive when there is economic
advantage, and repulsive for economic disadvantage\footnote{It does
  not matter here whether we consider the force to be a Newtonian
  deterministic force, or a probabilistic susceptibility for drifting
  closer, as in stochastic systems. We can think of a generalized
  energy-momentum tensor\cite{burgesscovariant}, for generalized agent `fields'.}.  The
formation of superagents thus comes about, for economic
reasons\cite{sirimace2007}, by the value of collaboration.  If the
promises are unconditional, superagents will be localized.  If they
are conditional, the clusters are ordered and may thus be distended or
even scattered.
\begin{itemize}
\item Agents, which make the same kinds of promises of same polarity, tend to repel one another.
\item Agents, which make complementary (binding) promises of opposite polarity
tend to attract one another.
\end{itemize}

Applied to the city problem, this suggests that the basic attraction
to condense the city out of a surrounding gas of agents comes from the
common supply promises, which are predominantly of positive sign, and
that all agents share; for the survival infrastructure. This is held
in check by the repulsion of agents making similar promises, except
where there are promises to cooperate.  One may expect structures as
follows:
\begin{itemize}
\item Clusters of professionals bound together by cooperation promises.
\item Chained transport agents, bound together by conditional promises.
\item Distributed competitors, perhaps clustering around shared infrastructure hubs, 
e.g. malls, districts.
\end{itemize}
The embedded spacetime structure of the city should be an equilibrium
configuration between the attraction to needs and desires, and the
repulsion from competition with similar skills.

In promise theory, a specialized role characterizes a pattern of agents that make
similar promises.  By specializing specific tasks to specific agents,
each agent can be more focused in learning and adapting, but acquires
an additional cost of cooperation proportional to some positive power
of the cluster size. Superagency collaboration is a {\em short range}
interaction (between interior promises)\cite{spacetime2}. It can form
{\em long range} strong interactions, with associated cost, if it has
the internal resources to support these. Limiting interactions to
short range leads to stability.  The long range interactions drive
superlinear effects, but also promote `chaos'.
\begin{figure}[ht]
\begin{center}
\includegraphics[width=8cm]{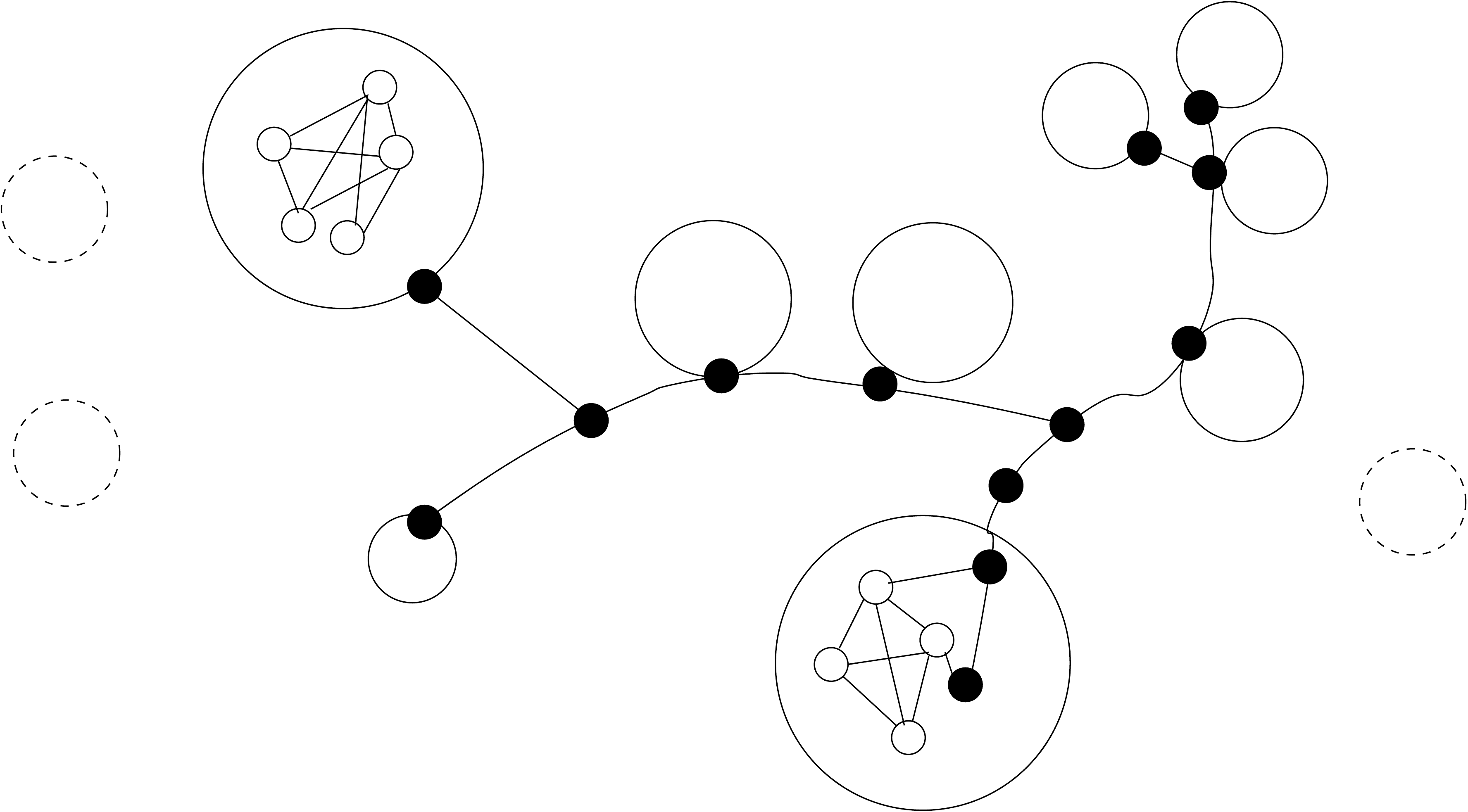}
\caption{\small The geometry of superagents may fill space in
different ways. Infrastructure that interconnects other agencies
is a superagent in its own right, involving linear or approximately
linear cooperation between member agents. Under preferential
attachment, agents $N_I$ tend to cluster around the infrastructure
agency, leaving a few $N_0$ padding out the spatial volume. The circles
around the subagents may be considered infrastructure binding
the agents together.\label{infragent}}
\end{center}
\end{figure}

The notions of attraction and repulsion are wired into our
imaginations in terms of spatial concepts. Even without an embedding
spacetime, we can speak of agent affinities, like the interactions
described in molecular chemistry, where spacetime plays no real role.
With a physical volume to embed a graph of promise-keeping
interactions, geometry ties range to distance, but in a virtual
network (which includes transport of messages by intermediate
carrier), short range interactions can also be disseminated over a
longer effective range, by adding cost or latency (such as in
telecommunications).

\subsection{Promise networks that percolate}\label{whypercolate}

In order to justify Metcalfe's law, there needs to be an average level
of interaction that spans the complete graph and propagates value.
This doesn't necessarily require a single promise type to dominate the
entire graph, because it is the value graph, not the promise graph
that needs to link up all the nodes. However, from the previous
section, we would expect basic infrastructure to dominate. The main
requirement for this is the presence of a common (or at least
interchangeable) currency between all the nodes.

Specialized exterior service promises naturally lead to small
molecular clusters of component `atoms' (superagents), that make
specific interior promises.  They seldom span large areas, because
promises act like short range interactions (which is also why
superagents can be considered quasi-atomic black box agents, see
figure \ref{percolate}), so they do not easily bring about percolation
of value in an economic zone, like a city. 
The promises, which are ubiquitous, are associated with the survival of
agents, and relies on the most general kind of infrastructure in
systems: power, food, air, water, etc. These are likely responsible
for the interconnection of the many smaller microcosms of value creation
(small businesses in cities, and microservices in IT) to bring about
a unified community with its economies of scale.

If we let $N_\tau$ be the number of agents that consume a promise of
type $\tau$, then we expect the class of $\tau$ related to survival to
be of the order $N_\text{survival} \simeq N_I$, in the meaning of the
city model. For all other types, $N_\text{other} \ll N_I$.
However, we'll see in section \ref{recursion} that long range
interactions are also needed to explain the scaling exponents for
cities. The size of the effective network is not therefore given by
the adjacency matrix of the underlying infrastructure network, but
rather by the typed promise graph.

It is useful to recall the definition of a promise network (see
\cite{spacetime2}).
\begin{definition}[Promise adjacency matrix]
The directed graph adjacency matrix which records a link if there is a promise
of any type $\tau$, and body $b_{ij}(\tau)$ between the labelled agents.
\beq
\Pi_{ij}^{(\tau)} = \left.
\begin{array}{cc}
1& ~ {\rm iff}~ A_i \promise{b_{ij}(\tau)} A_j, \\
0&
\end{array}\right\}~~~~\forall\; b_{ij}(\tau) \not=\emptyset
\eeq
\end{definition}
The adjacency is the effective topology of the spacetime network, as
far as the agents are concerned.
The link-occupancy of this matrix, for a given promise type, is a linear
sum whose value is generally much
lower than that of the total possible mesh of interactions. Thus, for any promise type $\tau$,
\beq 
\sum_{i,j=1}^{N_I} \Pi_{ij}^{(\tau)} = N_\tau(N_\tau-1)  \ll N_I^2,
\eeq
Note that an agent can make a promise
to itself too, so the upper limit could be written $N_I^2$.
The value-percolating connectivity or degree of a node 
\beq
\Pi_{ij} &=& \sum_\tau  \Pi_{ij}^{(\tau)},\\
k_i &\simeq& \sum_j \Pi_{ij}.
\eeq
We can also write this in terms of the direct valuation of the promises,
in terms of the actual matrix of promises $\pi_{ij}$\cite{spacetime2}:
\beq
k_i &\simeq& \sum_j v_C\left(\pi_{ij}\right).
\eeq
where $v_C$ is the value of the promise as
calibrated and assessed by a common central agency (see appendix
\ref{valuation}).

Agents can keep multiple promises and multiple types of promise `simultaneously' over
a given timescale $T$, by multiplexing their time at a rate that is much
faster, i.e. $t_\text{multi}\gg 1/T$ to avoid the queueing instability. On the assumption
of sufficiently sparse packing:
\beq 
\sum_\tau \sum_{i,j=1}^{N} \Pi_{ij}^{(\tau)} \le N(N-1).
\eeq

For the economic output of a promise network, we care more about the
assessments of which promises were kept than the number of promises
that were made (see appendix \ref{valuation}).  Each agent assesses
promises individually, and they may not agree. However, to compare to
city statistics, we may assume that an statistical bureau agency has
been appointed by the city to calibrate these assessments
$\alpha_\text{official}(\pi_\tau)$ according to a single scale.
Promise-keeping is an average over time.  Provided the sum time to
keep a promise, for all $\tau$, for each agent, is much less than each
time interval of the assessments, we can reduce $\alpha(\pi)$ to a
frequency `probability'.  Another way of saying this is: provided the
cost of keeping the promises is less than the budget of each agent.

These estimates are maximal.
The size of a functional cluster is not really related to any of these graphs,
because there are semantic constraints. Specific functional behaviour, in a
single promise type, is a strict
limitation, which leads to very sparse subgraphs. To gauge an average measure
of the total economic impact of all functional interactions, we have
to assume:
\begin{itemize}
\item The functions are successful in driving an economy.
\item The density of implicit interactions is quite high, else a given output
$Y$ will not be represented by an average mesh density.
\item There are some long range interactions that make the partially connected
graph totally connected on average, even if only at a low level. The survival
promises probably fulfill this role.
\end{itemize}
In reality, a city or community might be partitioned into quite independent regions,
leading to a modular reducible form\cite{graphpaper}.
If one imagines
the specific network, which delivers output $Y$, it may be some maximally quadratic polynomial
of $N_I$, related to the structure function of the network, but it may also be
significantly less than this. What will tend to lead to percolation of
interactions, which bind in a mesh, is the existence of long range, pre-conditional
promises, i.e. dependencies. Then, the sum will be some polynomial of $N_I$, such that
\beq 
\sum_\tau \sum_{i,j=1}^{N} \Pi_{ij}^{(\rm Y)} \simeq c_1 N + c_2 N^2.
\eeq
If $i,j$ run over all the individual agents within city limits, then
these matrices are sparse and fragmented for each $\tau$. 
Suppose we assume that there is no dominant infrastructure, only
small clusters of voluntary cooperation, as in \cite{siri3}. Then, 
the aggregate graph for a city
output $Y$, would need to have sufficient random cooperative 
connectivity to form a process that
generates output algorithmically. The density can be estimated,
if we assume that, for promise type $\tau$, an agent has an average valency
$\cal V$, then we would need
\beq
\sum_{\tau=1}^{\tau_\text{max}}\,\sum_{ij}^{\cal V_\tau}\alpha_C\left( \Pi_{ij}^{(\tau)}\right) \simeq N{\cal V}\tau_\text{max} \le N^2.
\eeq
Moreover, since the diversity of promise types is unlikely to be
greater than the population it stems from, one estimates $\frac{N}{\cal V}\le
\tau_\text{max}\le N$, which does not seem realistic. Taking these
estimates into account, it seems most
likely that a few types of promise relationships dominate the
connectivity and value creation, i.e.  the basic `survival' (food,
water, sanitation, communications) promises, and the specialized
outputs contribute relatively little to the total output, except when
directly dependent on the survival promises. This is a further
suggestion that the attraction to urban life stems from
core interconnection infrastructure, rather than from independent diversity.

\subsection{Scaling of roles and interactions}

Consider now the role of dependence in keeping promises as a determinant for scaling.
\subsubsection{Linear consumption by independent agents}

For promises that are requirements for survival, every agent more or less
deterministically depends on some source infrastructure agent $S$, the number
of promises is one to one:
\beq
\text{Number of consumed} = \sum_{i=1}^{N_I-1} \Pi_{iS}^{(\tau)} \simeq N_I-1.
\eeq
If the source is distributed over several agents, then this is still true:
\beq
\text{Number of consumed} = \sum_{s=1}^S \sum_{i=1}^{N_I-1} \Pi_{is}^{(\tau)} \simeq N_I-1.
\eeq
Thus, the number of promises needed to supply this demand is also
proportional to $N_I$:
\beq
N_\text{infra} \equiv N_\tau \simeq N_I.
\eeq

\subsubsection{Sublinear economy of scale, and spacetime involvement}

When agents interact through links, on a small scale, the
chemistry of their interactions may be based on simple counting.
Normally, in promise theory, we count by agent or by link. However, as
the numbers of converging links become so great that counting is
impractical, there is no way to liken a process to a simple Poisson
arrival queue, and we resort to flow counting based on density arguments.
Let's now show that this is equivalent to volume of the infrastructure $V_I$
in \cite{bettencourt1}:

Consider a number of agents $N_\text{infra}$ who provide
infrastructure (gas stations, supermarket, etc) for a number of
clients $N_\text{client}$, which in turn offer a service conditionally,
based on the dependence. 
\beq
\pi_\text{infra}:~~~ A_\text{infra} &\promise{+\text{infra}\#{\cal V}}& \{ A_\text{client}\}\\
\{ A_\text{client}\} &\promise{-\text{infra}}& A_\text{infra}\\
\{ A_\text{client}\} &\promise{+\text{service}|\text{infra}}& A_\text{infra} .
\eeq
Suppose that each infrastructure agent $A_\text{infra}$ can promise to service $\cal V$ clients
simultaneously; then, using a simple valency argument, we have a detailed balance equation
for the interactions at steady state:
\beq
\alpha_+\,N_\text{infra}\,{\cal V} \ge \alpha_- \, N_\text{client}.
\eeq
Thus for simple counting of distinguishable agents, we may estimate the number of
infrastructure agents needed to support a number of clients:
\beq
N_\text{infra} \ge \underbrace{\left(\frac{\alpha_-}{\alpha_+}\right)\frac{1}{{\cal V}}}_{\text{intrinsic}} \times N_\text{client}.
\eeq
where ${\alpha_-}/{\alpha_+}$ may be interpreted as the affinity for the service,
or the reciprocal compressibility. This scales linearly with the number of clients
in the catchment area of the infrastructure. Moreover, there is no way, in this
detailed formulation that we can count otherwise. The only economy of scale in
this arrangement is the standard linear multiplexing result for the marshalling of $\cal V$
queues into a single queue with $\cal V$ servers, noted in section \ref{uslsec}.

\begin{figure}[ht]
\begin{center}
\includegraphics[width=7cm]{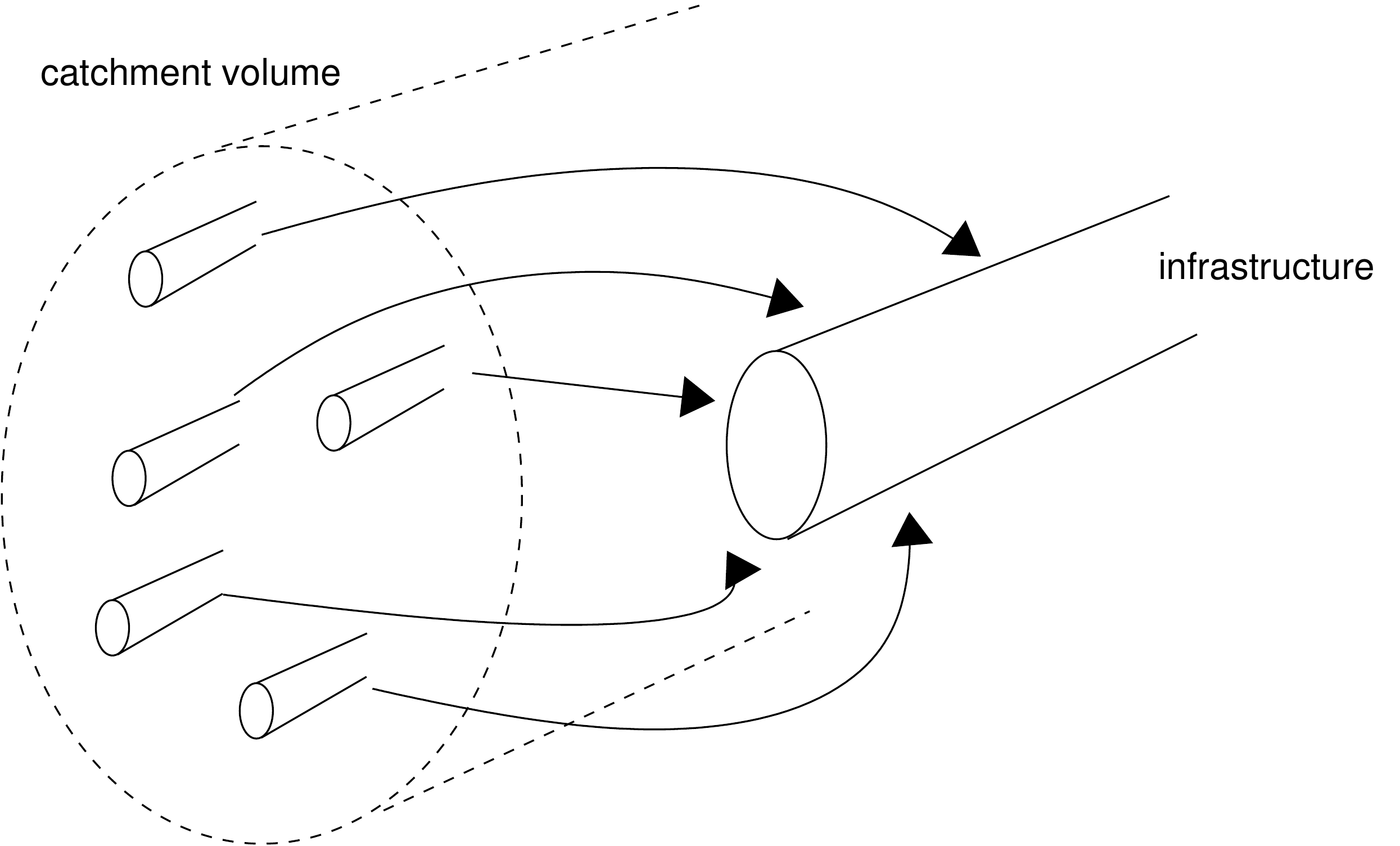}
\caption{\small How spacetime involvement compresses serialized agent links into an effective
flow of fixed cross section.\label{squeeze}}
\end{center}
\end{figure}

However, if we now ask how to count the number of clients that can be fed into
a single infrastructure agent, in a spatial volume, with dimensional
multiplexing, then the best estimate is to serialize the counting, as before:
\beq
 N_\text{client} &=& \left(\frac{V_\text{catchment}}{N_\text{users}}\right)^\frac{1}{D} \,C(D)\times N_\text{users},
\eeq
where we imagine a catchment volume $V_\text{catchment}$, containing any number of
agents $N_\text{users}$ who are interested in the infrastructure service, and
we serialize them along a tube of constant cross section $C(D)$ (see figure \ref{squeeze}). 
Although these numbers only apply to a small mesoscopic volume of space, in a
homogenous city, this will apply to the entire city, so we are
justified in taking
\beq
N_\text{use} \simeq N_I,
\eeq
which, combined with an equation of state for the volume, reproduces the earlier result
\beq
N_\text{infra} \simeq V^{1/D} N^{1-1/D}.
\eeq
In this argument, it is clear that only the active agents $N_I$ play a role in the
counting, and process flow, hence this also justifies why we can assume $N_I \rightarrow N$
in \cite{bettencourt1}.

\subsubsection{Deriving Metcalfe's law from promise networks: the importance of conditional dependency}\label{metcalfe}

A key assumption in the scaling argument of city outputs in
\cite{bettencourt1}, is what is Metcalfe's
law\cite{metcalfe,refmetcalfe1,refmetcalfe2}, which proposes that the
value of a network scales like the square of the total number of
nodes, i.e.  value is generated by the number of possible links
between agents\footnote{Metcalfe's law does not refer to the
  cost/value of physical connectivity, which (once again) can be much
  sparser than a mesh at low utilization. Indeed, that is important,
  else the net profit approaches zero. Rather, it refers to a
  correlation of agents' promised activities, linking their behaviours
  and generating value by interaction.}.  This has received some
empirical support in \cite{facebooktencent}, but has also been
criticized in \cite{refmetcalfe1,refmetcalfe2}.  More importantly,
interaction value generation is not the same as output.  Agents can
also produce wealth without interacting, if they have all the
prerequisites (short range interactions).  In section \ref{recursion},
we'll see that long range interaction forms the basis for one kind of
value creation, but not the only one.

Promise theory predicts that links represent value in the following
way. Consider then the sum of all impartial promise valuations by third party $C$.
If we assume that all agents assess the value of interactions as strictly positive, then:
\beq
\text{Mean value} = \sum_{\tau}\sum_{i,j=1}^{N_I} v_C(\pi_{ij}^\tau) \le c \langle \alpha_i\alpha_j\rangle  N_I(N_I-1)
\eeq
where $N_I = \max_\tau(\dim(\tau))$ (see appendix \ref{valuation} about promise valuations).
In spite of the quadratic appearance from the result, this
is a linear sum, so it acts automatically as a linear averaging
measure.  Also, for any given specialization $\tau$, the filling fraction of
the promise network is likely low; thus, a key assumption is that,
when properly documented, agent's specialized promises in fact depend on many others
{\em conditionally}, forming a wide reaching network of progressively
weak coupling. Conditional promises propagate the range of value
interaction\cite{faults,promisebook}. This is the ecosystem effect.

The weakness of coupling is not a problem provided the city is
reasonably homogeneous in density over the timescales of
the ensemble parameters, because of the assumption of
overall sparseness.  If we define an effective density
for the network, which describes some probable average level $\rho
\in[0,1]$ of `intercourse' between agents (any kind of sustained
relationship), then it is fair to write the value of a network of
promises: 
\beq 
\text{Mean value} = \sum_{\tau}\sum_{i,j=1}^{N_I} v_C(\pi_{ij}^\tau) =
c\, \rho\; N_I(N_I-1).  
\eeq 
provided the total density of promises
forms an SCC of order $N_I$ members.  This value can be distorted from
the quadratic form by significant inhomogeneity.  Now, for most
cities, $N_I \gg 1$ and $\rho, \alpha_i \ll 1$, so for strictly positive
value interactions: 
\beq 
v \simeq c\, \rho N_I^2.  
\eeq 
This is Metcalfe's law. It depends on the assumption of strictly
positive value (i.e. no non-profitable interactions), and sufficient
density of promises to involve everyone in the city who belongs to the
infrastructure.  Why is this plausible, when most specialization leads
to modularity?  One reason is that modularity is only a separation of
scales, not an elimination of dependency: dependencies form an
ecosystem. Nearest neighbours might hold the greatest semantic
importance to a given function, but this reductionist viewpoint is not
independently sustained without the eigenstability of the entire
web\cite{graphpaper}.

\subsection{Conditional dependency as an explanation for multiple superlinearity classes}\label{recursion}

We can note briefly why certain system processes (or occupations, in a
city) scale differently. In a specialization society, singular
individuals or agencies rarely have all the prerequisites to complete
their work without assistance.  They need to collaborate and depend on
others. Thus other agents take on the role of effective infrastructure
for one another. It is the accessibility of this dependency on one
another that throttles output, and can modulate scaling behaviour.

The argument for superlinear scaling in cities in \cite{bettencourt1}
uses interactions, with arbitrary (unknown mean field) range that can
span the city, to model economic output.  However, from a promise
theory perspective, it is reasonable to ask the question: should we
consider the output to be a fraction of $N^2$, representing the
maximum output due to interactions (as in Metcalf's law), or should we
consider the output to come directly from the agents, as a fraction of
$N$, as in Amdahl's law.  This offers an explanation for the anomalous
superlinear exponents in the data for \cite{bettencourt2}.  The
superlinear scaling was initially associated with `innovation'
activities\footnote{In \cite{arbesman1} an explanation based on
  spanning tree branching processes was postulated, but was not
  credible.}.  However, the promise theory shows how one does not need
to invoke a process of innovation to explain the scaling.  To drive
the long range cohesion of the whole community network, specialists
come to depend on specialized services (e.g. patents depend on
lawyers).  This leads to a number of cases:
\begin{enumerate}
\item {\bf Interaction scaling}: as proposed in \cite{bettencourt1}, for interactive
value creation.
\beq
A_\text{Lab}&\promise{+\text{patent}}& A_\text{observer}\\
A_\text{Lab} &\promise{\pm\text{interact}}&A_\text{services}
\eeq
Patent agencies are interacting at arbitrary range with a significant
fraction the total promise graph, as a part of the ecosystem.
\beq
Y \simeq Y_0 \left(\frac{v}{V_I}\right) N^2 \rightarrow  N^{1+\delta} \simeq N^\frac{7}{6} = N^{1.16}.
\eeq
The amplified value relies on the interplay between long range mixing, 
and short range isolation.

\item {\bf Scarce agent scaling}: 
skilled specialist experts' output is proportional to the number of skilled
agents, since their queue is sparse, and not filled by a wide
volume of demand. However, the same economy of scale applies to their services
when they are depended on, as `infrastructure', by others.
\beq
Y \simeq Y_0 \left(\frac{v}{V_I}\right) N \rightarrow  N^{\delta} \simeq N^\frac{1}{6} = N^{0.16}.
\eeq
\item {\bf Interaction promises with a scarce dependency}: 
such as in the case of a service that depends
on a source of agents to fulfill a dependency. e.g. patents can only be produced by labs
that depend on the outputs of specialized
R\&D employees and lawyers, working in private relationships, or in secrecy.
The expression in (\ref{depend}) assumes a promise configuration like that of the
assisted promise law\cite{promisebook}, with a main output based on a number of
agents that provide input. The dependencies produce raw output, and the 
`lab' agency collates and represents the collaborative mixing, e.g.
\beq
A_\text{Lab}&\promise{+\text{patent}|\text{research,legal}}& A_\text{observer}\\
A_\text{Lab} &\promise{\pm\text{interact}}&A_\text{services}\\
A_\text{Lab} &\promise{-\text{research}}&A_\text{staff}\\
A_\text{Lab} &\promise{-\text{legal}}&A_\text{lawyer}\\
A_\text{staff} &\promise{+\text{research}}& A_\text{Lab}\\
A_\text{lawyer}&\promise{+\text{legal}}& A_\text{Lab}
\eeq
More generically, with two stages in the process of promise keeping, each experiencing scaling (see figure \ref{SRD}),
\begin{figure}[ht]
\begin{center}
\includegraphics[width=7cm]{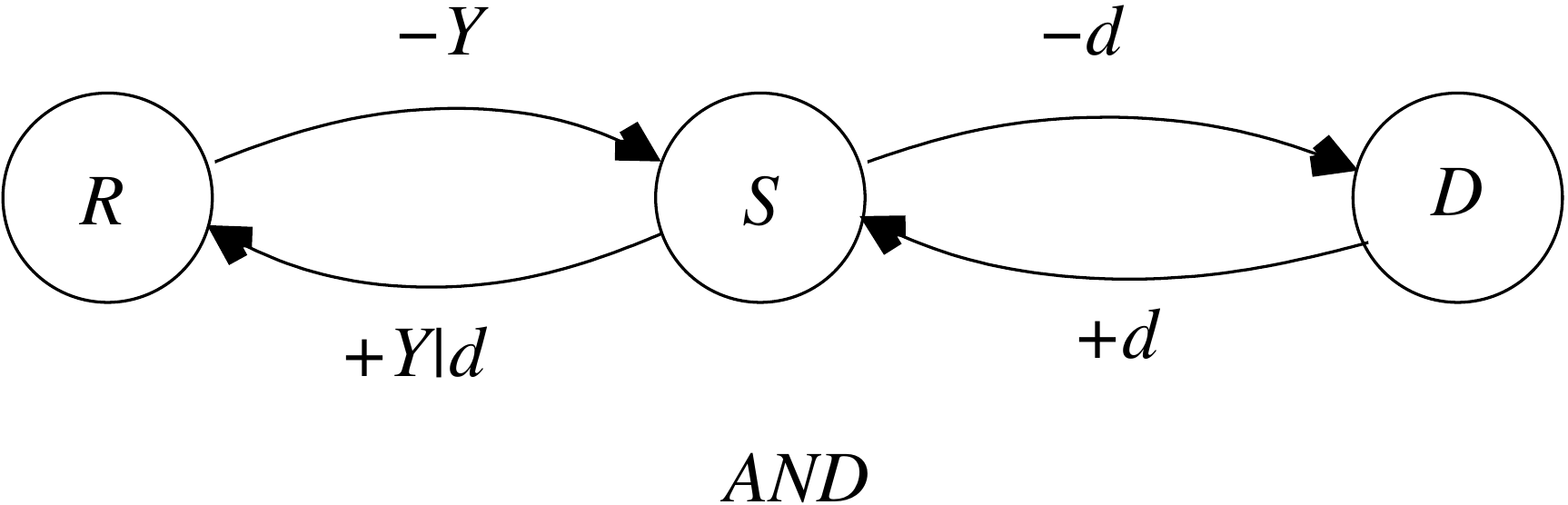}
\caption{\small A two stage (long range) dependency has two economies
of scale, when fed by a spacetime workflow. The probability of promises kept
is multiplicative, like the logical `AND' of the promises.\label{SRD}}
\end{center}
\end{figure}
\beq
S &\promise{+Y|d}& R\\
R &\promise{-Y}& S\\
S &\promise{-d}& D\\
D &\promise{+d}& S
\eeq
the total process picks up two `economies of scale': the delivery of $Y$ conditionally
AND the delivery of the conditional dependence.
\beq
Y \simeq Y_0 \left(\frac{v}{V_I}\right) N^2 \times\left(\frac{v}{V_\text{depend}}\right) N \rightarrow  N^{1+2\delta}\simeq N^\frac{4}{3} = N^{1.33}.\label{depend}
\eeq
where $D=2$ is used for the numerical values. These values accord better with the
cited data in \cite{bettencourt2}, and tie in with the story about queueing.

What characterizes this interaction is the high level of
specialization required to fulfill the dependencies. If the network is
sparse, this is more difficult than if it is dense and diverse. This
is the specialization gamble. With specialization comes individual
efficiency, but also risk of instability by disconnection from key
dependencies\cite{tainter}.

They are a throttle on the
process, because their absence could stop it altogether. Hence, we are
justified in using the product `AND' for combining the probably
values in (\ref{depend}).
\begin{figure}[ht]
\begin{center}
\includegraphics[width=7cm]{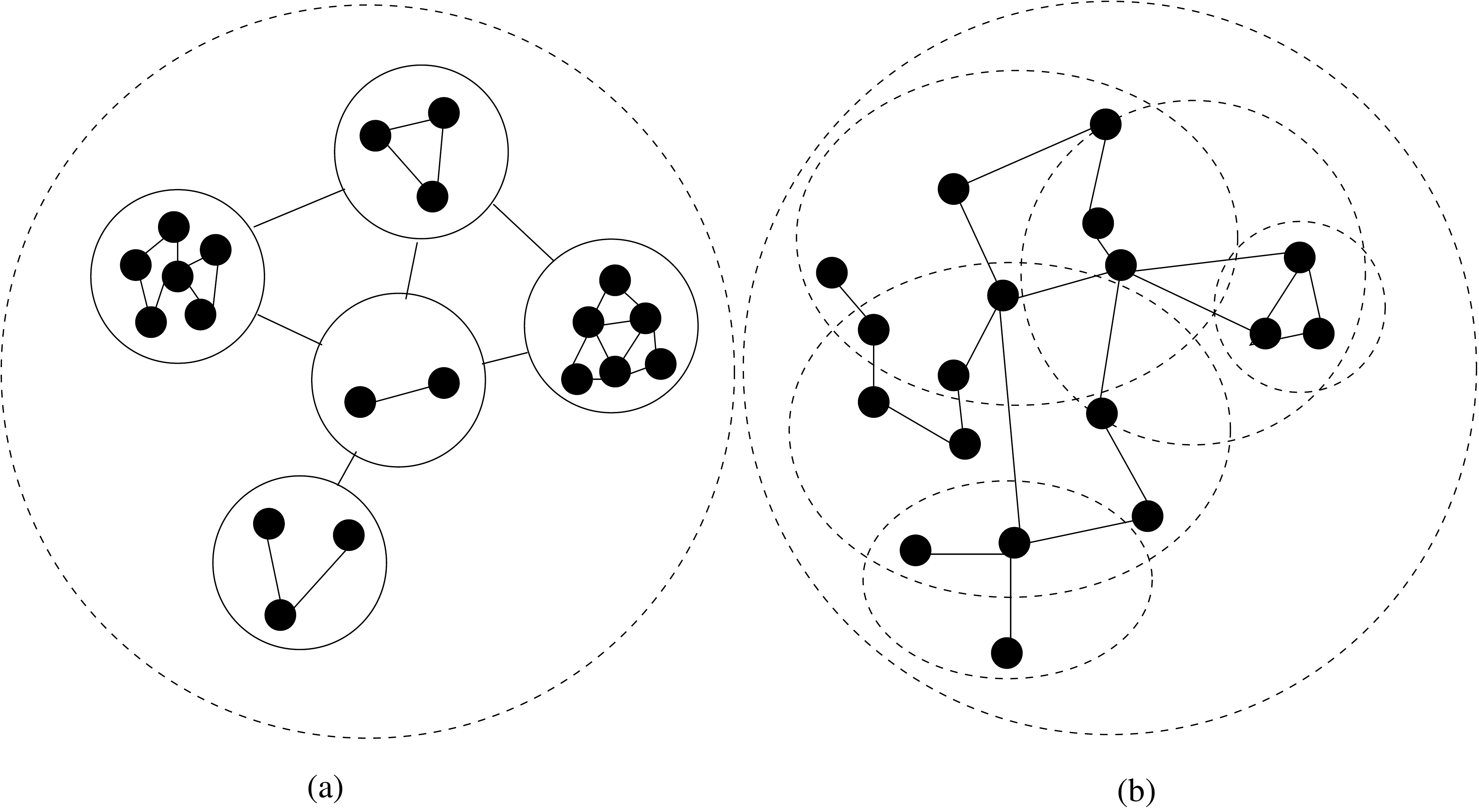}
\caption{\small Structural recursion in an ecosystem is not like a branching process of containers (a), but rather the agents overlap with other regions of the same network to access their
virtual functions. Thus their outputs are not concealed as interior substructure, but
exposed as part of the flat internetwork (b). The result is that
a second order recursion picks up a second economy of scale, in turn increasing the
superlinearity of the derived output.\label{hierarchy}}
\end{center}
\end{figure}
This is not a hierarchical system interaction, because the services are
not necessarily hidden from the long range dynamics by internal
components of the superagent `lab' (see figure \ref{hierarchy} (b)).
But in organizational theory, one normally assumes that all
organizations are hierarchically organized (see figure \ref{hierarchy}
(a)).

\item {\bf Recursive promise dependency}.  Let's consider what happens
  when the ecosystem network is based on a hierarchy of interaction
  ranges, i.e. promises are made recursively in fully protected
  shells. The agency produces a service using full community
  infrastructure, but also has some specialist dependency contained
  entirely within (see figure \ref{hierarchy} (a)).

\beq
A_\text{company} \promise{+A_\text{specialist}\promise{+\text{solution}} A_\text{client}}  A_\text{client}
\eeq
would be viewed as a recursive operation on the infrastructure, and 
the economics of scale would apply to both times the (different) infrastructures
were used.
\beq
Y_\text{patent} &=& \frac{N_I^2}{V_\text{patent}}\\
V_\text{patent} &=& g_\text{R\&D}\left(\frac{V_\text{R\&D}}{N_\text{R\&D}}\right)^\frac{1}{D} N_\text{R\&D}\\
V_\text{R\&D} &=& \left(\frac{V}{N_I}\right)^\frac{1}{D}
\eeq
Substituting for $V_\text{R\&D}/N_\text{R\&D}$ from the last expression into the former,
\beq
V_\text{patent} &=& g_\text{R\&D}\left(\frac{V}{N_I}\right)^\frac{1}{D^2} N_\text{R\&D}
\eeq
Inserting this into the output expression
\beq
Y_\text{patent} &\simeq& N^{1+\frac{1}{D^2}-\frac{1}{D(D+1)}} \simeq N^\frac{13}{12} \simeq N^{1.08}.
\eeq
This value is almost linear, which is what we might expect on a self-contained
specialization, since the outside world would not be able to tell the difference
between a single agent and a single superagent.

The value is also smaller than case (1) above, not larger, so short
range hierarchical scaling cannot explain the anomalously large
exponents measured in cities. On the other hand, there are some
smaller exponents in this range. It is interesting to examine these
measures from the perspective of the promises represented, and their
range in the embedding space.
\end{enumerate}

There is a simple prediction here: long range interaction via dependency seems to
increase output superlinearity, by compounding ecomomies of scale,
i.e. dependency brings strong long range coupling and activates a
larger amount of the $N^2$ mesh.
A similar effect can be obtained artificially in \cite{bettencourt1},
by slightly increasing the Hausdorff dimension of the infrastructure
$H > 1$. This does would correspond to a more pervasive generic
infrastructure network, which is an opaque explanation at best. It
seems unclear how to justify it.

Short range dependency is basically invisible at larger scales.  This
observation might help to explain superlinear seen in technological
contexts too, through coordination\cite{gunther3}, but we have to be
careful not to mix together effects that come about due to higher
dimensionality, with other mechanisms for increasing the utilization
of dependent resources.

\subsection{Service discovery of dependencies in a network}\label{howinnov}

The ability for agents to discover and match with other agents, that
make complementary promises, is the basis of functional scaling, and
the semantics of cooperation and innovation. It depends on either
physical or virtual mobility of the agents.  Kinetically, agents may
follow a random walk, as in ballistic discovery. A second possibility
is that cheap intermediaries perform the discovery of specialized
roles\footnote{Electrons play this role in molecular chemistry, or
  telecommunications in the human realm.}.  We can say that two agents
are either
\begin{itemize}
\item Physically close.
\item Virtually close.
\end{itemize}
Promise theory also suggests that they may be close in two ways: dynamically close or
semantically close (such as when related meanings are similar). The
former depends on the length scales of the system (e.g. city) and its
structure. The latter can be assumed approximately independent of
these scales, because the carriers are very light, cheap, or fast (or,
in the case of semantic distance, purely cognitive). If
the cost of discovery can be neglected, the cost equation
is different: collaboration can be cheaper, and the
value of being in close proximity for a particular specialization is
reduced\footnote{A dependency does not just have to be discovered, but
  also maintained in a persistent relationship, which accumulates cost
  over time.}.

Directories, maps, and indices\cite{spacetime2} are the keys for
agents to virtualize discovery of dependencies, and locate one another
without physical search in spacetime.  Telephone directories map
coordinate addresses to names. Yellow pages map coordinates to
specializations.  Similar specializations are grouped.  Shopping malls
and industrial estates act as physical directories, where clients can
expect to find services in a small volume. Directories may be
discovered themselves, or formed by voluntary registration.  The value
of new bindings overcomes the tendency for similar specializations to
repel one another: similar agents may be attracted implicitly
(covalently) by the intermediate attraction to clients. Apart from
predicting the importance of directories for smart cities, and
organizations, this also predicts that the availability of directory
information could affects the productivity of a city as a function of
size. This effect might not be clearly visible in the ensemble data, since
we would need to compare cities at the same value of $N$.

The scaling estimates of the city are based on infrastructure where
physical motion of the population is based on the cost of traversing
some fraction of the length of the city. We can repeat the output
calculation to neglect this cost, as is the case in services that
do not require physical transport.

\begin{itemize}
\item {\bf Physical interaction} (transport/mobility): 
people move around using transport infrastructure
to experience their environment.
There is a promise for people to observe their surroundings,
for something related to subject $\tau$,
and this promise is kept fractionally $\alpha_\tau \in [0,1]$ during their walk.

Let the linear range of the agent $A_i$ be some dimensionless fraction
per unit time $rT_{\rm explore}$ of the size of the city $V^{1/D}$,
where $r$ is the speed in units of city size\footnote{If the person's
  path is detailed, one could include the Hausdorff dimension of the
  path and use $v_i^{H/D}$ as the range, as Bettencourt suggests. I'll
  ignore this for now, as humans do not tend to move in fractal paths,
  as his data suggest.}. If the density of impulses per unit length of
city region $\cal I$ is assumed constant relative to the transport
rate (because this is the basis of commerce, i.e. what the city is
trying to optimize for people's finite time), then the number of
impulses $I_\tau$ of type $\tau$, experienced on such a walk, may be
written: 
\beq 
I_\tau \propto r_i\; T_{\rm explore}\; V^{1/D}\; {\cal I}\; \alpha_\tau 
\eeq
where $\alpha_\tau$ is the probability that the person or agent will
be receptive to impulses in its environment that are relevant to
promises of type $\tau$.

Although there is room for inhomogeneous variations in the city
regions, in the transport rate $r$, and the density of offerings $\cal
I$, this will not change the average scaling argument much, as long as
$N$ is large.  I make the assumption here that the density of
experiences $\cal I$ is constant, even though the density of people is
related to the city size.  This is because the size of a city is
constrained by the time rather than the distance (and we are
suppressing explicit time).

The range will be some fraction of the size of the city, available
by transport infrastructure $V^{1/D}$.
The cost of physically fishing for ideas thus takes the form
\beq
C \simeq c_Y N_I V^{\frac{1}{D}},
\eeq
in agreement with the work model of \cite{bettencourt1}. This applies for physical city
interactions, and leads to the same output scaling expression in \cite{bettencourt1}.
\beq
Y^+_Y &\simeq& N^{\frac{2D+1}{D(1+D)}} N_I^{\frac{D^2-D}{D(D+1)}},\\
&\simeq& N_I^{\frac{7}{6}}\left(1+\frac{N_0}{N_I}\right)^{\frac{5}{6}}.
\eeq

\item {\bf Virtual interaction} (teleport/messaging): people are immobile and send
  messages to one another, watch entertainment, browse, read,
  talk, etc. These activities occupy an increasing
  amount of the time spent by people, not least because it can easily
  be interleaved with work time. The rate is no longer related to the
  size of the city, nor is there any obvious boundary to what can be
  discovered online (since the range of the Internet is even more
  diverse than a city)\footnote{Because telecommunications networks
    are global, it does not make sense to relate their cost to the size
    of the city (though this depends on exactly how we model the
    costs), so the cost depends more on its usage than on its extent.
    We simply assume that it exists and has sufficient capacity for
    the $N_I$ connected residents.}. In this case, the impulses are
  more likely to be related to availability of the fountain itself
  (e.g. `bandwidth' $B$) multiplied the time spent.
\beq 
I_\tau \propto B\; T_{\rm explore}\; {\cal I}\; \alpha_\tau 
\eeq
Discovery of information is the main issue. Before search engines,
there were only directories such as white pages (by person) and yellow
pages (by promise type). However delivery of what is discovered might still
involve spatial constraints, e.g. locating a new car online does not
allow it to be teleported to the buyer's location. However, 3d printing
technology might change this, for a class of problems,  soon.

Here it is not the locations that matter, but the rate at which
impulses are absorbed. Once again, this is constant. When friends,
books, or movies are communicating ideas to us, this happens at a rate
that depends only on how quickly we can get hold of a stream.  How
users discover locations online, or by telephone is a separate
question. Directories\cite{spacetime2}, advertisements, and chance all play a role
here.
The cost of fishing for ideas is thus now
  independent of the city size. For a community of multiple super-agents,
the analogous expression is:
\beq
C \simeq c N_I B T_{\rm explore}\label{net}.
\eeq
If we imagine a community with no other infrastructure except
its telecommunications network, and substitute (\ref{net}) into
the detailed balance equation:
\beq
g_Y\left(\frac{v_Y}{V}\right) N_I^2 \ge c N_I B  T_{\rm explore}.
\eeq
Following through the calculation for the yield estimate
identically, we find the scaling is no longer superlinear ($D=2, H=1$)
\beq
Y \simeq N^{\frac{H}{D}}N_I^{\left(1-\frac{H}{D}\right)} \simeq N_I^\frac{1}{2}\left(1+\frac{N_0}{N_I}\right)^{-\frac{1}{2}}.
\eeq
This simple result reflects the intuition that, if we neglect the
`universal cost' of telecommunications from the community accounting,
then the value generated as a result of collaborative processes is
proportional only to the fraction of participants who span the
diameter of the city or community. This reproduces the well-known result for
mobile ad hoc networking (MANET)\cite{burgesslongversion,burgessim2003}.
\end{itemize}
The rate of output based on trawling of ideas and gestation in closed workgroups will be
\beq
I^{\rm work}_\tau = N_D \left( c_{\rm phys} I_\tau^\text{phys} +c_{\rm virt} I_\tau^\text{virt}  \right)
\eeq
and for the entire city of $N_W$ workplaces:
\beq
I^{\rm city} = \sum_\tau I^{\rm work}_\tau\simeq N_W \overline I^{\rm work}.
\eeq
The physical channel agrees with Bettencourt's expression for mixing volume
in (\ref{vie}), where $N_I=N_DN_W$, and the virtual channel accords with
mobile ad hoc networks. It would interesting to examine communities
physically remote from services to see how these predictions match reality.

%%%%%%%%%%%%%%%%%%%%%%%%%%%%%%%%%%%%%%%%%%%%%%%%%%%%%%%%%%%
\section{Remarks on technological infrastructure and collaborative networks}
%%%%%%%%%%%%%%%%%%%%%%%%%%%%%%%%%%%%%%%%%%%%%%%%%%%%%%%%%%

Cities are just one form of smart adaptive space, for which we now
have a new and fascinating insight in the form of statistical scaling
data.  Remarkably little data are available for computer
installations, software development, or smart warehouses, since these
reside in the private sector; nonetheless, there may well be insights
to gain from studying the relationship between theory and practice,
where we can. Some results may have sufficient dynamical similarity to
other cases to infer valuable lessons. Although there are qualitative
differences between biological organisms and cities, the main features
that makes computer systems different from cities are the size and
timescales involved.  Datacentres are still tiny compared to cities
(in terms of active agents $N$).  Structural changes take place on the
order of seconds rather than weeks, and the rates increase as
technology advances (see figure \ref{timescales}). The chief lesson we
can derive from cities, which might be applied to other smart
infrastructure, is the involvement of spacetime relationships in
counting, at large $N$.

In \cite{spacetime1,spacetime2}, a generalized abstraction of
functional spaces was developed, starting from `atomic' irreducible
considerations. By developing the city scaling theory in this
framework, one could hope to bridge the gap between disciplines, and
promote future analysis of the effect of changing costs and
technology. In information technology, dynamical infrastructure, known
as cloud computing, offering co-located shared compute, storage and
caching, as well as providing facilities for community software
development, shared repository models, and finally the pervasive
Internet of Things, represent both present and upcoming
challenges for infrastructure modelling.

\begin{figure}
\begin{tabular}{|c|c|c|}
\hline
\sc IT infrastructure & \sc Common & \sc City\\
\hline\hline
         & $N \simeq 10-10^7$ &       \\
\hline
&  Functional modules &\\
\hline
&  Long and short range interactions &\\
\hline
$< 10^{-3}s$ & \sc Network $\Delta t_{\rm infra}$   & $> 10^3s$\\
\hline
$10^{-6}s - 1s$ & \sc Agents $\Delta t_{\rm interaction}$ & $1s - 10^{4}s$ \\
\hline
Seconds & \sc Transport time & Hours\\
\hline
software    & \sc Innovation & hardware \\
\hline
  code, services   & \sc Promises      &  goods, services    \\ 
\hline
code, memes, habits    & \sc Epidemic transmission & replicants, memes, habits\\
\hline
Membership & \sc Semantic edge & Membership\\
\hline
Latency threshold & \sc Dynamic edge & Density threshold\\
\hline
Servers, storage & \sc Tenancy unit & Homes, offices, storage\\
\hline
Containers, hosts, private nets & \sc Partitioning & cubicles, rooms, buildings\\
\hline
Process groups, clusters, datacentres & \sc Super-agencies & cubicles, rooms, buildings\\
\hline
$H = 1 - N$ & \sc Trawl path dimension $H$ & H = 1-2\\
\hline
$1-3$ & \sc Embedding dimension & $D = 2-3$\\
\hline
\end{tabular}
\caption{Comparing cities with IT infrastructure. \label{timescales}}
\end{figure}

\subsection{Online communities versus infrastructure clusters}

We must be clear about the difference between online communities and physical
networks.  It has not escaped the notice of
\cite{bettencourt3} that universality of cities implies network
scaling in technology-assisted communities, and some data
concerning online communities has been examined with interesting and
large superlinear behaviour:
\begin{center}
\begin{tabular}{c|c|c}
\sc Measure & Average $\beta$ & Source\\
\hline
DNS hosts  & $1.28\pm 0.06$ & Internet\\
Total web pages & $2.03\pm 0.1$ & Internet \\
Active web pages & $1.68 \pm 0.1$& Internet\\
\hline
Contributors & $1.61\pm 0.1$ & Wikipedia \\
External links & $1.59 \pm 0.2$ & Wikipedia \\
Internal links & $1.21\pm 0.02$ & Wikipedia \\
\end{tabular}
\end{center}
Insufficient details are provided to suggest an explanation for the
numbers\footnote{\cite{ygchen1} has made some comments on the
  distribution of community sizes for physical communities.}; however,
a promise theory approach like the one begun here, may easily play a
role in understanding the structural dependences.

Online communities are sociologically interesting, but more practical,
from a engineering viewpoint, is to understand how scaling behaviour
modulates productivity in smart infrastructure, that mixes physical
and virtual mobility. Such `smart spaces' enact a kind of computation
to optimize their conditions, and even exhibit some algorithmic qualities.
Users interact with hosted services, which are
themselves a community of (software) agents, acting as proxies for
human intentions. This client interaction behaves like a long-range
weak coupling: an autonomous `gas' of visitors\footnote{In
  \cite{burgessIJMPC,burgessPRE1}, I showed how statistical behaviours
  in computers behave like a gas at equilibrium, with periodic
  boundary conditions.}, and malls and directories act as catalysts
for value generating interactions.  Applications and companies, on the
other hand, consume unique and shared resources as residents of the
infrastructure. They have superagent boundaries around company and
functional concerns, which limit internal processes to short-range
(and typically strong coupling) interactions, in the manner of a
tenancy\cite{spacetime2}.
Some brief remarks below concern what one might expect to study and
find in software agents that are proxies for human intent, and are enhanced with
significant automation.

\subsection{The productivity of agents in human-machine systems}

Productive output, in agent communities, is driven either by output
from agents working independently, in parallel, or from the mesh of
interactions between them. It is throttled by the contention for
shared resources (serialization or queueing) and the high cost of long
range coherence (i.e. equilibration of state, or mutual calibration). These are the main
features captured in Gunther's Universal Scaling\cite{gunther1}.

Automation allows individual agents to generate much greater outputs,
without the cost of cooperation, so unequal automation might skew the
measured performance outputs in empirical studies, making it difficult
to compare cities and other systems\footnote{Some notes to this effect
  were remarked in \cite{bettencourt4}.}.  When forming a statistical
ensemble of systems (cities, organisms, or cloud infrastructure), we
have to be sure to compare similar systems. A fundamentally different
technology base would undermine the universality of data for
individual data points, and could influence the interpretation of the
scaling.

As pointed out in section \ref{howinnov}, the impact of using
messenger channels, like telecommunications rather the physical
mixing, is that the process of information and service `discovery' is
fundamentally different, altering the costs by removing the
imprint of physical scales from the output scaling. Caching becomes
a crucial enabler here, because it converts quadratic $N^2$ costs in to
linear $N$ outputs from single agents (thanks to smart behaviour).

For technology infrastructure, the analogue of discovery by wandering
around a town browsing store fronts, in the high street or shopping
mall, is browsing a database, or directory (e.g.  yellow pages, or
online shopping catalogue).  Information spread by rumour and
reputation, in a pre-telecommunications communities, are supplemented
by targeted information about recommendations (e.g.  Amazon, Google,
and Facebook ads and search).

Productivity is not only increased by automation. The cost of
transporting and equilibrating goods and services remains. In
information technology, larger and larger amounts of data are now
stored in physical and data warehouses. Distribution channels, for
products and services are mirrored by technologies for data
replication, such as Paxos\cite{paxos} and Raft\cite{raft}. These are
now being adopted widely, based on the sense that data equilibrium is
important to avoid the inconsistency of `many worlds' viewpoints, when
interacting with distributed systems.  However, the scaling of
equilibration software is necessarily poor, since the cost scales
superlinearly (typically like $N^2$), to maintain consistent
states\cite{gunther3}.
Resource `latency' or response time is a key issue, when dependencies
are not local. This too can be improved by caching. The cost of
transporting data and materials from remote suppliers, either to a city is
like the cost of reaching across the planet to rent a virtual machine
platform.  Ultimately, no community wants to rely on a fragile remote
resource.  A cheap non-scalable solution is better than an expensive
long term one. This has driven the centralization of large cloud
installations, where economies of scale can be argued locally,
as transport costs and latencies are the responsibility of the clients.

Caching and replication are strategies that decouple long-range
dependences, and restore agent autonomy.  It is cheap to cache and
replicate many technology functions today, offering a kind of `smart
behaviour' like a brain to keep interactions local.  The mobility of
small data is high, as transferring small data is cheap. The mobility
of computation has traditionally been low, as computers were large and
cumbersome machinery.  But that is reversing, thanks to encapsulation
methods (superagent technologies), like machine virtualization and
so-called `containerization' of software.  Moreover, there is
computational processing capacity everywhere today, including on
smartphones in our pockets, whereas the sheer size of data being
collected is growing to impractical levels for transportation.
Computation is progressing from being a shared resource, to a
ubiquitous agent capability.

\subsection{Separation of long and short range interactions (modules)}

Limiting unnecessary mixing due to long-range or strong-coupling
interactions is essential for establishing modular functionality in
networks.  Modularity promotes specialization, and the quiet isolation
for the gestation of lengthy processes like learning and innovation.
It makes economic sense, provided the modules do not themselves become
bottlenecks. Where humans are coupled strongly, the Dunbar limits for
human valency\cite{dunbar2} intrude on scalable design, even with the help of
automation.

What does this tell us about cloud computing, microservice
communities, and the Internet of Things? Suppose the economy of scale
were only of the order of 85\%, as in cities, this would not a huge
incentive to centralize, if local resources were available too.  The
economy would have to be offset against effects like latency and long
range dependence, equilibration, etc.  As our interest in data from
embedded sources increases, and we equip environments with smart
ecosystems\cite{certainty}, the idea that cloud computing will be
localized in large datacentres seems unrealistic.  Latency costs
suggest a greater delocalization of workforce to eliminate long range
interaction.

Managing specialization, or separation of concerns, is a
human-technology issue that we are only just starting to grapple with
at scale\cite{workspaces}.  Modularity has long been a part of system
doctrine, but the evolution of so-called `microservices', or small,
specialized software services, is now being motivated empirically, by
the limitation of Dunbar valency for human agents\cite{dunbar1,dunbar2}.  Breaking a system
into many small specialist parts, each associated with a different
human owner, incurs a new cost of service interaction, but this cost
may be paid cheaply with automation to alleviate a more expensive
human burden: the hard limit of what a human brain can cope with.  The
density of information services in modern society is now huge, and the
complexity of interactions is significantly more than humans can
manage unassisted.  Technological agents can handle the greater
numbers of interactions, but only if they are sufficiently simple that
a human collaboration can understand how to design the promises they
should keep.  Coordination services for replicating siloed resource
clusters are being developed, based on the experiences of large
industrial providers like Google\cite{kubernetes}.  These currently
rely on long-range data equilibration methods, which is serious
throttle on their performance, and must become worse when spacetime
dimension plays a role.

Scaling issues pose the question: at what point should systems
(cities, datacentres, communities) break up into decentralized
regions?  Cloud datacentres grew up from the economies of scale that
can be achieved through specialized expertise in infrastructure
management. We have already witnessed cloud datacentres multiply like
power stations, placed at strategic geographic locations, to cover
distances. The next logical step is ubiquitous infrastructure. At this
point, the economic advantages of physical clustering (indeed cities
themselves) may disappear altogether, leaving only the value of
centralized meeting places, clubs, conferences, etc for human contact.

\subsection{Semantic separation versus dynamical separation (silos)}

Silos that encapsulate specializations are isolated semantic
functions.  The appearance of silos in human organizations is often
thought of as a negative phenomenon, because it represents an exclusion
of outside interests.  However, it also has a necessary and positive
effect, implying a separation of short range interactions.  

This is not the same as separation of dynamical scales.  Voluntary
partitioning to mitigate contention has clear advantages, if resources
permit.  Conversely, confusing semantic separation with physical
separation may have dynamical consequences. An excellent example of
this may be seen from town planning: for a while `garden cities' were
a trend, designed with the aim of tidy separation of functions,
separated by green spaces. The principle backfired, however, e.g. in
Brasilia\cite{brasilia}, were different city functions were so
distributed that residents have to travel considerable distances to
reach town facilities. This led to severe traffic congestion.

When the cost of collaboration between partitioned agencies
grows\cite{tainter}, business networks can degenerate and become
non-viable, as a larger part of the infrastructure becomes
non-contributing. This is a lesson for microservices.  IT architects
are starting to realize this now, and are opting for ubiquitous
`hyperconverged' architectures, i.e. a return to total package servers
to reduce latency and congestion. Sometimes regions form by themselves, from dynamical and
semantic principles, with partitions based on language, business,
geographic centrality, eigenvector centrality\cite{archipelago}, etc.

\subsection{The mobility of agents in human-machine communities}

In the past, sparse machine resources were immobile, and inputs were
brought to the machine for processing. It was cheaper to move inputs
to machinery.  Today, mobile devices with significant processing
capability are commonplace.  and encompass computers and manufacturing
(aka 3d printing). Particularly, in information infrastructure, the
ubiquity of stationary and mobile processing capacity reduces the need
for data to be sent over large distances. Conversely, the amount of
data expected to come from domestic and industrial sources will grow
massively, decreasing relative mobility of data. Mobility of
processing resources is now a crucial issue, and is enabled by
containment wrapper technologies that form intentional superagent
modules around specific functional roles.

The ubiquity and miniaturization of information technology suggests
that the role of space dimension may well be a temporary phenomenon,
at least for local interactions.  Even with ubiquitous information
infrastructure, there will still be some role for large datacentres,
particularly in the realm of storage archives, since disaster
redundancy is one of the critical issues. Similarly, `data gravity' proposes
moving the smaller resource to the larger resource, e.g.  move
computational power to large data instead of assuming that
computational power has to be at a fixed location and transporting
data.  

\subsection{Scaling of voluntary cooperation versus imposition}

Garbage collection, sewage, and drainage services are amongst the most
important dependencies for a system.  They are sinks for pushed
output: scaling results for these services would be interesting
indeed. One would expect them to be more susceptible to flash floods
and long tail behaviours than the corresponding supply networks, as
they can only fail catastrophically when a threshold is
reached\cite{faults}. It would be very interesting to know how the
converse of `push' methods, or imposition services, scale with city
size: pull on demand, voluntary cooperation are popping up in society
to replace mass broadcasting, e.g. video on demand, replacing
television broadcasts.

\subsection{The rate of change in the system and utilization}

The perception of time has an interesting relationship with long range
order. Time is a local concept, as both physicists and distributed
systems specialists know only too well.  Without long range data
synchronization, clocks may not tell the same time, and not just human
clocks, but the pulses that drive and clock processes\footnote{Each
  change in an agent is a tick of the network clock, and each
  partitioned network has its own sense of time, and time only moves
  forward according to that clock when something happens (this is
  essentially Einsteinian relativity\cite{spacetime1}).  }.  For
example, imagine an orchestra of musicians. The orchestra can play
without a conductor is everyone has enough to do all the time, but
when every instrument plays a sparse role in the whole, coordination
is difficult as the agents are mostly dormant.  Multiplexing allows
better utilization of a network. When all agents are busy, i.e.
utilization is high, process time passes at a faster rate, relative to
its surroundings, and coordination becomes easier.  Busy agents are
ready to receive new inputs and respond with services, suggesting that
a highly utilized system could be more efficient, provided that it
lies below the queueing instability.

At high speed, the world is dominated by time, and is essentially one
dimensional.  In a world dominated by space there is
multi-dimensionality, and volumes can quickly become relevant in a
continuum limit, and time is suppressed by equilibration.  The higher
dimensionality promotes an increased average connectivity for the
promise graph, $k(N) \simeq N^\delta(D)$, that depends on the
dimension of the embedding space $D$.  In the complex interaction
between space {\em and} time, all possibilities are on the table.

\section{Summary}

Universality and scaling are powerful notions in science.  Having data about the
scaling of functional processes, at large and small $N$, offers an invaluable
insight into what we can expect of technological systems at scale, and
their increasing intrusion into human society.
Understanding social sciences in terms of laws, analogous to physical
law, is an area where progress has been made over the past century.
Understanding such patterns in `smart' admixtures of humans and
technology seems an even more relevant challenge
today\cite{burgessbook2}.  An obvious question becomes: is there
something that can be transferred from the study of cities to other
network and community systems, e.g.
\begin{itemize}
\item Online user communities---human communities in a virtual space.
\item Humans-software communities---interacting processes in a virtual space.
\item Microservice communities---collaborating agents in a virtual space, interacting with humans.
\item Cloud computing---a shared infrastructure community of contending processes.
\end{itemize}
Datacentres and software systems share a lot of similarities with
cities, but there are differences too. Information infrastructure is
largely one dimensional, in most locations. In datacentres
infrastructure is only just in the process of becoming two and even three
dimensional.  It is also quite inhomogeneous, and one needs to know
when and where dynamical similarities might be exploited to argue
dynamical similarity\cite{certainty}.  As $N$
becomes large, issues of space and time become much more entwined with
the more common one-dimensional algorithmic behaviours generally
studied in computer science. 

Universality reveals emergent laws, on broad scales. However, a fuller
understand of systems, whether human cities, smart cities, computers,
or any other human structure, is only achieved by describing both
dynamics and semantics at micro- and mesoscopic scales.  Just as we cannot
understand medicine without understanding the functional roles of
structures inside organisms, so the functional organs in a city are
key to what it does. The universal scaling arguments for urban areas,
in \cite{bettencourt1, bettencourt2}, are exciting discoveries, as
they point to the involvement of spacetime in functional systems,
which is still poorly understood in information and computer science.

Datacentres are still tiny compared to cities, but the two are also
growing together thanks to the pervasive spread of the Internet.  By
applying promise theory to expose some short range interactions, we
find more possible exponents that match quite well with the anomalous
results in cite{bettencourt1} (see section \ref{recursion}).  The
immediate applications of this approach leads to
some basic observations:

\begin{itemize}
\item Value creation in communities comes from a mesh of promises
  whose outcomes funnelled, filtered and powered by serialized
  processes for supply and harvesting.

\item The sparse probabilistic utilization of shared infrastructure
  allows agents to interleave their efforts and achieve economies of
  scale. This keeps a fluctuating city in an approximately stable,
  steady state over short timescales, and admits limited long term
  growth.

\item Specialization of tasks into modulaar services allows
  systems to focus their time and capabilities without the cost of
  context switching. The strategy of specialization also brings
  fragility: if the cost of reconnecting the specialists grows too
  large, the community can fail to keep promises for essential
  functions\cite{tainter}.

\item Increasing autonomy of a system population, due to the
  availability of personal assistants, and localized capabilities
  (smartphones, 3d printers, etc) will undermine the need for
  transport, and the involvement of city scales in many processes.

\item Superlinear scaling results from a dependence on exterior
  specialist agencies. When remote dependencies are involved, staged
  economies of scale can accumulate bringing superlinear effects, for
  each remote dependence to the ensemble. If these external
  dependencies could be redistributed to a purely autonomous attribute
  of each agent (e.g.  when phone boxes are replaced by mobile
  phones), then scaling would at best be linear, and the artefact of
  superlinearity would disappear.

\item The ability to discover promised information, to mix it and
  select new combinations, is the basis of innovation and
  collaboration.  A highly discoverable ecosystem is `smart' in the
  sense that it can adapt and invent. Any space can, in principle, be
  made smart in this way, if its agencies make the necessary
  infrastructure promises.
\end{itemize}
There are two main lessons to take from these notes, that may be
surprising to technologists: (i) the behaviour of a system can involve
more than one spatial dimension, and (ii) we can measure and describe
just how smart and productive functional spaces actually are across
a range of scales from local to global.

\section*{Acknowledgments}

Many thanks to Luis Bettencourt for patiently reading and refining my
understanding of his model.  I'm grateful to Noah Brier for an
invitation to Transition in late September 2015, where I met Geoffrey
West and learned about scaling in cities.  This work could not have
happened but for that chance meeting in a remote city. Also thanks to
John D. Cook for carefully reading a draft.

\bibliographystyle{unsrt}
\bibliography{spacetime} 

\appendix
\section{The value of promises}\label{valuation}

The value of links in a network depends on the promises they make.
The value of a promise is a form of assessment\cite{promisebook} that any agent can make
independently. We write an assessment of whether a promise was kept
\beq
\alpha_i(A_i\promise{b}A_k) \in [0,1]
\eeq
to mean the assessment by agent $A_i$ that the promise from $A_j$ to $A_k$ was kept.
A valuation is an estimate of what a promise is worth to an agent. This may of may not
depend on the assessment of to what extent the promise is kept or not.
Every agent assesses on its own calibrated scale. If we want a common currency
valuation for all parties, this has to be calibrated by a single agent according to its
scale.

The interpretation of value is also an individual judgement that
relies on trust, and may be based on accounting of the assessments
over time (reputation)\cite{burgesstrust}.

\beq
\text{My reputation} \propto \sum_\text{you} \left(\text{you}\promise{-b}\text{me}\right)
\eeq
In words, my reputation is proportional to all the number of `you' who
(publicly) promise to accept my promised service.
Even unilateral promises may have some value:
\begin{center}
\begin{tabular}{c|c|c}
\sc Valuation by $X$ & \sc About promise & \sc Reason for value to $X$\\
\hline
  Me & $\text{me}\promise{+b}\text{you}$ & An reputation building investment\\
  Me & $\text{you}\promise{+b}\text{me}$ & A service that might help me\\
 You & $\text{me}\promise{+b}\text{you}$ & A service that might help you\\
 You & $\text{you}\promise{-b}\text{me}$ & You need the service now\\
\end{tabular}
\end{center}
Cooperative relationships are usually based on conditional assistance\cite{promisebook},
and take the form of a conditional equilibrium:
\beq
S \promise{+S | M} R\\
R \promise{+M | S} S\\
S \promise{-M} R\\
R \promise{-S} S.
\eeq
in words, $S$ promises $R$ a service, if it receives payment $M$; and
$R$ promises to pay $M$ if it receives service $S$.  
In a network without trust, this is a deadlock. But if any agent
trusts the other enough to go first, it is a cyclic generator of a
long term relationship. Such relationships imply lasting value, as
known from game theory (for a review see \cite{certainty}).  Both
agents also promise that they will take (-) what the other is offering
unconditionally. This is a signal of trust. Valuations are not necessarily rational to anyone but the agent that makes them,
and are unrelated to cost.

The economic value is that something is exchanged, which requires
a binding of both + and - promises.
\beq
S \promise{+S} R\\
R \promise{-S} S
\eeq
Both agents recognize the value of the other party, so the value 
exchanged is proportional their assessments that the promises were kept:
\beq
v_S \propto \alpha_S(S \promise{+S} R)\alpha_S(R \promise{-S} S)\\
v_R \propto \alpha_R(S \promise{+S} R)\alpha_R(R \promise{-S} S).
\eeq
In a community where such transfers are made often and between
arbitrary pairs of agents, standards of valuation are equilibrated, and
may be exchange in league with a calibration agency (e.g. a bank or government).
Thus, in a well-connected community, with a spanning infrastructure,
we may posit that the value of a one-way transfer is simply
\beq
v_C(\Pi^S_{ij}) = c_S \alpha_i\alpha_j, 
\eeq
where $c_S$ is the currency value of a perfect service relationship $S$, and
$\alpha_i$ is an impartial assessment of the probability with which $A_i$ will keep
its promise to give or receive $S$.

%%%%%%%%%%%%%%%%%%%%%%%%%%%%%%%%%%%%%%%%%%%%%%%%%%%%%%

\section{Defining the edge of a city or community by membership promises}\label{edge}

So far, we've skirted around the issue of what constitutes the city
limit, in the definition of a city. The size of the city plays a role
in the measurements, and the model in \cite{bettencourt1} treats the
city as an economically inflated balloon, with an edge, so we must
understand what constitutes the scope of the city or semantic space.
In fact the city is more like a club with membership than a balloon
with an edge.

Consider how the network we call a city is reached from beyond. How
does it communicate with the outside world? Cities have roads and
other infrastructure links in and out of the city (James Blish novels
notwithstanding). Should these links be treated as if they were part
of the same infrastructure mesh as the city itself? If so, where does
the city start and end? The use of dynamical scaling arguments for
everything else suggests that there might be a dynamical `network'
answer to this question of boundary conditions, but this is not the case.

\begin{figure}[ht]
\begin{center}
\includegraphics[width=4cm]{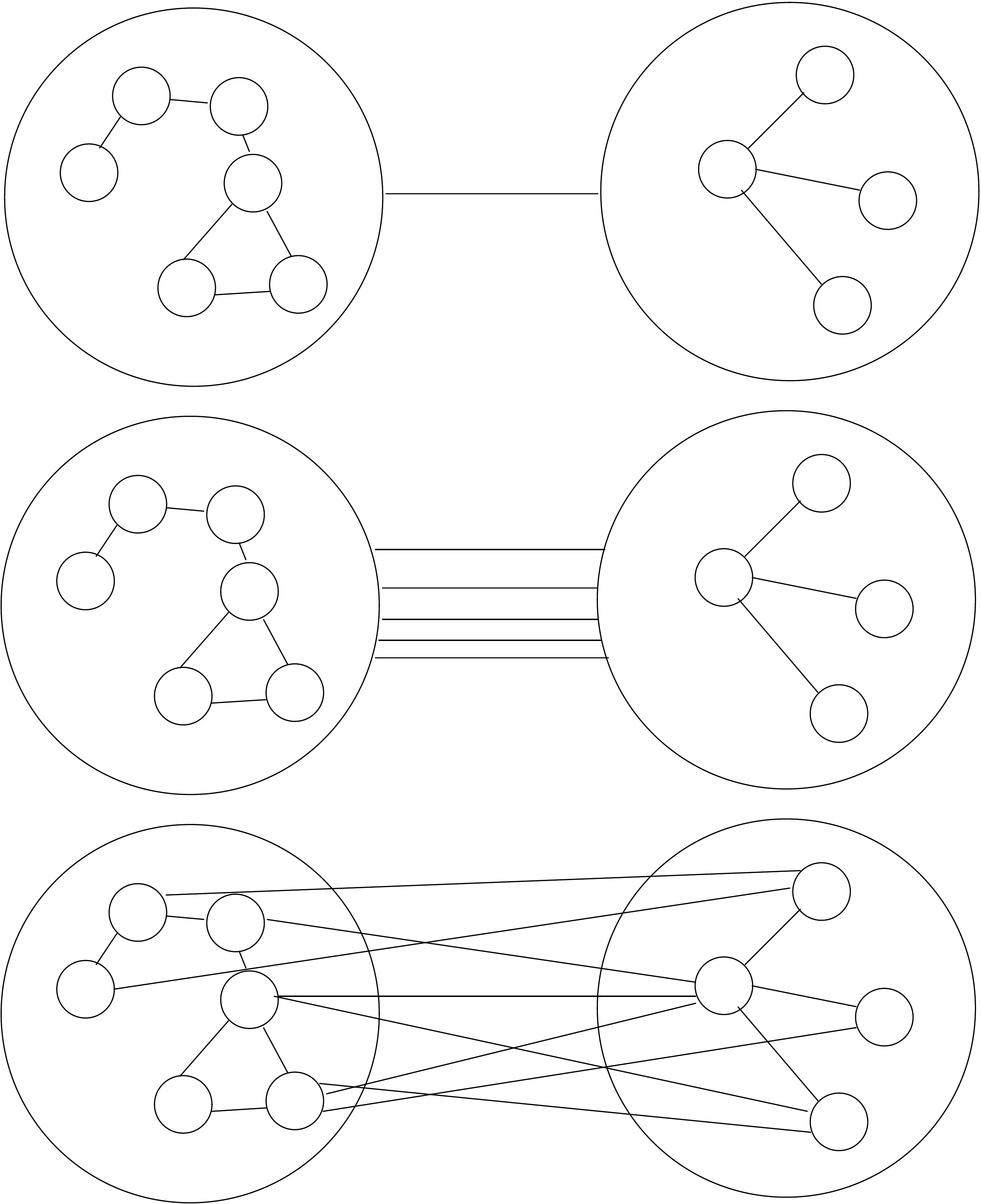}
\caption{\small A simple thought experiment: when do two cities merge
  into a single one? As we add more an more link capacity, when does
the scaling change from $N_1^2 + N_2^2$ to $(N_1+N_2)^2$? Providing the links are
not saturated, it takes only a single link. So where what does the
edge mean?
  \label{dumbell}}
\end{center}
\end{figure}

Consider a simple thought experiment, in which we start with two
separate cities and join them together by increasing the density of
connections (see figure \ref{dumbell}).
As soon as a city is connected to an outside network, this extends the
internal network. As long as the `external' network has the capacity to support
the aggregate level of traffic from the population $N_I$, then it is
no different from the internal infrastructure network, and we have simply
extended the boundary of the city.

One proposal might be to look for the presence of a discontinuity in
the network capacity, like a threshold event horizon, at which the
response time for a network interaction changed: \beq R_{\rm external}
\gg R_{\rm internal} \eeq But this doesn't quite make sense either.
Some processes are fast and some are slow, even in an urban centre (a
letter posted may take longer to arrive than an office worker takes to
process and even reply to it, while a person can take the train across
town in half that time)\footnote{An impartial approach, based on
  actual network topology, would be to use the `Archipelago method',
  to defining regions of network eigenvector centrality. Using a hill
  climbing to define natural regions that are seeded on very central
  nodes leads to well defined regions\cite{archipelago}. The question
  is whether these have any semantic significance. There are two ways
  to do it: either based on the shared infrastructure network, or on
  the virtual business networks that describe the outputs $Y$, by
  defining an effective adjacency matrix based on promise bindings.}.
An ecosystem has a broad mixture of timescales. Only by trying to separating
weakly coupled agents can we compare relative timescales for modular components.
Moreover, the entirely local gestation or production time is usually the limiting
time factor in the economic processes considered here, not the transport
or delivery times.

While physically plausible, this this is not how cities are defined.
They are defined semantically, by labelling of community membership.
The simple explanation\cite{sirimace2007,promisebook,spacetime1} is
that a city is defined to be that collection of agents that mutually
promise to be members of the city, and that are accepted as such by
the city authorities. In practice, the population must register as
residents, and they receive promises of services (including tax
collection). One assumes that the transient population of any city is
a small correction to this.

The autonomy of any observer counts in making the judgement of
community boundary.  Each agent can (and will) judge independently
whether it considers itself a part of a region or not. This begs the
question how collected data define communities, and they really have
an edge or not.  Can the scaling laws be made to fit any
interconnected network?  The structural considerations in section
\ref{recursion} suggest that, with the right understanding of
functional structure, the universalities can be applied properly.
These issues have been highlighted in the empirical studies in
\cite{Arcaute20140745,bettencourt4}, where it is found that the
scaling is distorted by picking the `wrong boundary' in urban regions.

As we try to apply the ideas to similar networks, such as IT
infrastructure and online communities, the role of the physical city
as an entity becomes unimportant. It is rather the community
that resides and interacts within it that plays the major role of
mixing. The details of the infrastructure play a role, but the
universality lies in the notion of a community.

\section{The phase of a community: mobility of agents, and interaction catalysts}

Agents exist a priori in an unbound state, effectively a gas phase,
mixing with their changing surroundings.  By promising constrained
cooperation, they can voluntarily become part of a solid phase,
interacting only with fixed neighbours.  Mobility of promising agents,
in the gas phase, remains important in all functional systems.

There are two kinds of networks in a city:
\begin{itemize}
\item Supply or delivery networks which are one-way flows
from source to destination. These are mainly branching
processes, but may also have simple redundancy.

\item Collaboration networks with two-way interactions between
communicating agents. The agents can be people, machines, companies,
etc.

\end{itemize}
Collaborative networks are based in interactions. Chemistry
demonstrates that such networks can be realized in two ways: either by
using fast messengers (electrons) between slow fixed molecules (like
covalence), or by moving faster molecules between slower interaction
regions (like ionic), and we may arrange the same methods in larger
systems.
\begin{itemize}
\item People at fixed locations can use telephone or Internet to send
  messages (solid state agents).
\item People travel between locations, carrying messages with them (agents as a gas).
\end{itemize}
These methods may be treated as different kinds of network in the model.
Promise theory is a chemistry for generalized semantics bindings.

The model proposed by \cite{bettencourt1} is close to the appearance
of a kinetic theory, but the city is not a gas with random motions in
this model. It's phase is not defined, because the physical
realization of the network is not defined. The movement of the
population could responsible for forming links between agents, i.e.
transport via the infrastructure networks (like taxis and subway); or
intermediate messenger technologies could be responsible (such as post and
Internet). The phase could play a role in the scaling of the infrastructure
network in general, constraining its degrees of freedom and range.
A solid phase limits the effective dimensionality $D$.

Modern cities comprise a fixed infrastructure in a mainly `solid
state', while the agents are sometimes bound in a solid state, and
sometimes free as a gas.  People move around the city in the subway
like water, but increasingly they use messages, like `covalently'
bonded work-molecules rather than transport pipes.  Now that
distance is less costly, the ease and speed of networking is
reflected in the lower density of modern cities, versus older cities
where high density enabled ease of meeting, in a kind of primordial
soup of mixed intercourse.

Cities are not the only functional networks, of course. Any community
could be completely mobile, with no fixed address, such as online
communities, coming together only in meeting places that play the role
of catalysts. This makes the mixing of skills and promises more
efficient\footnote{Curiously, it is also believed in biology that the
  cooking of food is what made humans an efficient species, supplying
  energy to fuel our large brains.}. Catalysts for bringing agents
together, like social meeting places play an increasingly important
role. Open source software is one of the important outputs of the
modern world, which happens completely outside cities\footnote{Github
  and other version control repository virtual code libraries function
  as catalytic meeting places.}.  In biology, the infrastructure
networks are in a fluid state, with functional agents (cells)
transported suspension, and their promises advertised by compatability
molecules and receptors on their surfaces, like exterior superagent
promises\cite{spacetime2}, exactly analogous to small businesses.  IT
networks are mainly solid, or quasi-solid, even when using mobile
devices, as the messages move between the agents much faster than the
agents move themselves so the motion of devices is negligible for
many purposes.

\section{Effective power law scaling from Amdahl's and Gunther's law}\label{notscin}

The Amdahl and Gunther scaling relations are workload scalings, not in
the usual form of universal scaling relations. Let's consider how we
might derive an approximate power law scaling relation from these.
If we want to know how the time fraction (speedup) scales for
as a function of the number of processors $N$, then we can compare $N$ with
$\gamma N$. Then, we can write, for $\gamma > 1$
\beq
\frac{T(\gamma N)}{T(N)} &=& \frac{\sigma + \frac{\pi}{\gamma N}}{\sigma + \frac{\pi}{N}}\\
&=&  1 + \frac{\left(\frac{1}{\gamma}-1\right)\frac{\pi}{\sigma n}}{1+\frac{\pi}{\sigma n}}.
\eeq
Let $\delta = \frac{D}{D+H}$, where $D = 1$ and $H = \frac{\pi}{\sigma n}$, then, approximating as a binomial expansion,
\beq
\frac{T(\gamma N)}{T(N)} &=& 1 - \left(\frac{\gamma -1}{\gamma}\right)\delta.\\
&\simeq&  \left(1-\left(\frac{\gamma -1}{\gamma}\right)\right)^{\delta} \ldots\\
&\simeq& \gamma^{-\delta}.
\eeq
Thus we have an approximate power law fit for large $N$ compared to $\pi/\sigma$, and we
can write
\beq
T(N) \simeq T_0 N^{-\delta}.
\eeq
i.e. there is a marginal relative economy of scale for small $N$, which decays to an 
essentially scale invariant constant result.
If we allow, like Gunther, for the presence of equilibration, or mesh coherence
effects, then we could use the form
\beq
T(N) = \sigma +  \frac{\pi}{\gamma N} + \kappa N,
\eeq
where $\kappa$ represents linear time take to poll each of the worker agents. This is
the case where replication and consistency are required. With this extra term, we have
\beq
\frac{T(\gamma N)}{T(N)} &=& \frac{\sigma + \frac{\pi}{\gamma N}+\kappa \gamma N}{\sigma + \frac{\pi}{N}+\kappa N}\\
&=&  1 + \frac{\left(\frac{1}{\gamma}-1\right)\frac{\pi}{\sigma}+ \left(\gamma-1\right) \frac{\kappa}{\sigma}N^2}
{1+\frac{\pi}{\sigma}+\frac{\kappa}{\sigma}N^2}.
\eeq
Now the behaviour doesn't separate cleanly, and there are two regimes of approximate power
law scaling, with something more messy in between. Using the same procedure as before,
we get an anomalous term for $\kappa\not=0$:
\beq
\frac{T(\gamma N)}{T(N)} &\simeq&  \left(1-\left(\frac{\gamma -1}{\gamma}\right)\right)^{\delta} + \frac{\kappa (\gamma-1) N^2}{\pi+\sigma N+\kappa N^2}\\
&\simeq& \gamma^{-\delta} ~~~~(\kappa \ll \sigma, N \text{small})\\
&\simeq& \gamma\; ~~~~(\kappa > 0, N \text{large})\\
\eeq
So for large $N$, with $\kappa > 0$, we have simply
\beq
T = T_0 N,
\eeq
i.e. the scaling cost becomes linearly worse with increasing size.

When we compare these results to a spacetime scaling, it will become apparent
that this takes the approximate form of a scaling law in a
one-dimensional spacetime $D=1$, with Hausdorff dimension $H =
\pi/\sigma n < 1$. This indicates that a serial workflow, with some
parallelism, is essentially a one dimensional problem, with some
fractal complexity in its trajectory due to
parallelism\footnote{Alternatively, if we think about the problem
  graph theoretically, we can also say that it behaves like a $D=N$
  dimensional space, and a trajectory with Hausdorff dimension $H =
  \pi/sigma$. In a graph, the node degree $k=N$ is the effective
  dimension of spacetime at the point\cite{spacetime1}.}.
Interestingly, as the parallelism increases, the duration of the
fractal dimensionality shrinks to nothing. Thus the large $N$ limit
for serial processing tends to squeeze the degrees of freedom in the system.

\end{document}